\documentclass[sn-standardnature]{sn-jnl}

\usepackage{multibib}
\newcites{m}{References}
\bibliographystylem{sn-standardnature}
\newcites{sup}{References (Continued)}
\bibliographystylesup{sn-standardnature}

\usepackage{graphicx}%
\usepackage{multirow}%
\usepackage{amsmath,amssymb,amsfonts}%
\usepackage{amsthm}%
\usepackage{mathrsfs}%
\usepackage[title]{appendix}%
\usepackage{xcolor}%
\usepackage{textcomp}%
\usepackage{manyfoot}%
\usepackage{booktabs}%
\usepackage{algorithm}%
\usepackage{algorithmicx}%
\usepackage{algpseudocode}%
\usepackage{listings}%
\usepackage{xspace}

\theoremstyle{thmstyleone}%
\newcommand{\NII}{{[N\,{\sc ii}]}}
\newcommand{\NIIs}{{[N\,{\sc ii}]\,}}

\newcommand{\NeIII}{{[Ne\,{\sc iii}]}}

\newcommand{\SII}{{[S\,{\sc ii}]}}
\newcommand{\SIIs}{{[S\,{\sc ii}]\,}}

\newcommand{\OIII}{{[O\,{\sc iii}]}}
\newcommand{\OIV}{{[O\,{\sc iv}]}}
\newcommand{\OIIIs}{{[O\,{\sc iii}]\,}}

\newcommand{\OII}{{[O\,{\sc ii}]}}

\newcommand{\HeII}{{He\,{\sc ii}\,}}

\newcommand{\Ha}{H$\alpha$}
\newcommand{\Has}{H$\alpha$\,}
\newcommand{\Hb}{H$\beta$}
\newcommand{\Hg}{H$\gamma$}
\newcommand{\Hbs}{H$\beta$\,}

\theoremstyle{thmstyletwo}%

\theoremstyle{thmstylethree}%

\begin{document}

\title[A dormant black hole in the young universe]{
A dormant overmassive black hole in the early Universe}


\author*[1,2]{\fnm{Ignas} \sur{Juod\v{z}balis}}\email{ij284@cam.ac.uk}

\author[1,2,3]{\fnm{Roberto} \sur{Maiolino}}
\equalcont{These authors contributed equally to this work.}

\author[1,2]{\fnm{William} \sur{M. Baker}}
\equalcont{These authors contributed equally to this work.}

\author[1,2]{\fnm{Sandro} \sur{Tacchella}}
\equalcont{These authors contributed equally to this work.}

\author[1,2]{\fnm{Jan} \sur{Scholtz}}
\equalcont{These authors contributed equally to this work.}

\author[1,2]{\fnm{Francesco} \sur{D'Eugenio}}
\equalcont{These authors contributed equally to this work.}

\author[1, 2]{\fnm{Joris} \sur{Witstok}}
\equalcont{These authors contributed equally to this work.}

\author[3,4,5,6]{\fnm{Raffaella} \sur{Schneider}}
\equalcont{These authors contributed equally to this work.}

\author[3,4,5]{\fnm{Alessandro} \sur{Trinca}}
\equalcont{These authors contributed equally to this work.}

\author[4,5]{\fnm{Rosa} \sur{Valiante}}
\equalcont{These authors contributed equally to this work.}

\author[7]{\fnm{Christa} \sur{DeCoursey}}
\equalcont{These authors contributed equally to this work.}

\author[8]{\fnm{Mirko} \sur{Curti}}

\author[9]{\fnm{Stefano} \sur{Carniani}}

\author[10]{\fnm{Jacopo} \sur{Chevallard}}

\author[11]{\fnm{Anna} \sur{de Graaff}}

\author[12]{\fnm{Santiago} \sur{Arribas}}

\author[13]{\fnm{Jake} \sur{S. Bennett}}

\author[1, 14, 15]{\fnm{Martin} \sur{A. Bourne}}

\author[10]{\fnm{Andrew} \sur{J.\ Bunker}}

\author[16]{\fnm{St\'ephane} \sur{Charlot}}

\author[1]{\fnm{Brian} \sur{Jiang}}

\author[1, 14, 17, 18]{\fnm{Sophie} \sur{Koudmani}}

\author[12]{\fnm{Michele} \sur{Perna}}

\author[19]{\fnm{Brant} \sur{Robertson}}

\author[1, 14]{\fnm{Debora} \sur{Sijacki}}

\author[1,2]{\fnm{Hannah} \sur{\"Ubler}}

\author[20]{\fnm{Christina} \sur{C.\ Williams}}

\author[21]{\fnm{Chris} \sur{Willott}}

\affil[1]{\orgdiv{Kavli Institute for Cosmology}, \orgname{University of Cambridge}, \orgaddress{\street{Madingley Road}, \city{Cambridge}, \postcode{CB3 OHA}, \country{UK}}}

\affil[2]{\orgdiv{Cavendish Laboratory - Astrophysics Group}, \orgname{University of Cambridge}, \orgaddress{\street{ 19 JJ Thomson Avenue}, \city{Cambridge}, \postcode{CB3 OHE}, \country{UK}}}

\affil[3]{\orgdiv{Dipartimento di Fisica}, \orgname{“Sapienza” Universit\`a di Roma}, \orgaddress{\street{  Piazzale Aldo Moro 2}, \city{Roma}, \postcode{00185}, \country{Italy}}}

\affil[4]{\orgdiv{Osservatorio Astronomico di Roma}, \orgname{
INAF}, \orgaddress{\street{  Via di Frascati 33}, \city{Monte Porzio Catone}, \postcode{00040}, \country{Italy}}}

\affil[5]{\orgdiv{INFN, Sezione Roma1}, \orgname{“Sapienza” Universit\`a di Roma}, \orgaddress{\street{  Piazzale Aldo Moro 2}, \city{Roma}, \postcode{00185}, \country{Italy}}}

\affil[6]{\orgdiv{Sapienza School for Advanced Studies}, \orgaddress{\street{  Viale Regina Elena 291}, \city{Roma}, \postcode{00161}, \country{Italy}}}

\affil[7]{\orgdiv{Steward Observatory}, \orgname{University of Arizona}, \orgaddress{\street{933 N. Cherry Avenue}, \city{Tucson}, \postcode{85721}, \state{Arizona} \country{USA}}}

\affil[8]{\orgname{European Southern Observatory}, \orgaddress{\street{Karl-Schwarzschild-Strasse 2}, \city{Garching}, \postcode{D-85748},  \country{Germany}}}

\affil[9]{\orgname{Scuola Normale Superiore}, \orgaddress{\street{Piazza dei Cavalieri 7}, \city{Pisa}, \postcode{I-56126}, \country{Italy}}}

\affil[10]{\orgdiv{Department of Physics}, \orgname{University of Oxford}, \orgaddress{\street{Denys Wilkinson Building, Keble Road}, \city{Oxford}, \postcode{OX1 3RH},  \country{UK}}}

\affil[11]{\orgname{Max-Planck-Institut f\"ur Astronomie}, \orgaddress{\street{K\"onigstuhl 17}, \city{Heidelberg}, \postcode{D-69117},  \country{Germany}}}

\affil[12]{\orgdiv{Centro de Astrobiolog\'ia (CAB)}, \orgname{CSIC–INTA}, \orgaddress{\street{Cra. de Ajalvir Km.~4}, \city{Torrej\'on de Ardoz, Madrid}, \postcode{28850},  \country{Spain}}}

\affil[13]{\orgdiv{Center for Astrophysics}, \orgname{Harvard University}, \orgaddress{\street{60 Garden Street}, \city{Cambridge}, \postcode{02138}, \state{Massachusetts},  \country{USA}}}

\affil[14]{\orgdiv{Institute of Astronomy}, \orgname{University of Cambridge}, \orgaddress{\street{Madingley Road}, \city{Cambridge}, \postcode{CB3 0HA}, \country{UK}}}

\affil[15]{\orgdiv{Centre for Astrophysics Research, Department of Physics, Astronomy and Mathematics}, \orgname{University of Hertfordshire}, \orgaddress{\street{College Lane}, \city{Hatfield}, \postcode{AL10 9AB}, \country{UK}}}

\affil[16]{\orgdiv{Institut d’Astrophysique de Paris}, \orgname{Sorbonne Universit\'e, CNRS, UMR 7095}, \orgaddress{\street{98 bis bd Arago}, \city{ Paris}, \postcode{F-75014}, \country{France}}}

\affil[17]{\orgdiv{St Catharine's College}, \orgname{University of Cambridge}, \orgaddress{\street{Trumpington Street}, \city{Cambridge}, \postcode{CB2 1RL}, \country{UK}}}

\affil[18]{\orgdiv{Center for Computational Astrophysics}, \orgname{Flatiron Institute}, \orgaddress{\street{162 5$^{th}$ Avenue}, \city{New York}, \postcode{10010}, \country{USA}}}

\affil[19]{\orgdiv{Department of Astronomy and Astrophysics}, \orgname{University of California}, \orgaddress{\street{1156 High Street}, \city{Santa Cruz}, \postcode{96054}, \state{California},  \country{USA}}}

\affil[20]{\orgname{NSF’s National Optical-Infrared Astronomy Research Laboratory}, \orgaddress{\street{950 North Cherry Avenue}, \city{Tucson}, \postcode{85719}, \state{Arizona} \country{USA}}}

\affil[21]{\orgname{NRC Herzberg}, \orgaddress{\street{5071 West Saanich Rd}, \city{Victoria}, \postcode{BC V9E 2E7},  \country{Canada}}}


\abstract{Recent observations have found a large number of supermassive black holes already in place in the first few hundred million years after Big Bang, many of which appear overmassive relative to their host galaxy stellar mass when compared with local relation \citem{Harikane_AGN, Maiolino_AGN, Matthee2023, Ubler2023, Goulding2023_AGN, Kokorev2023_AGN,Ubler23a, Furtak2023_AGN,Maiolino24_GN-z11}. Several different models have proposed to explain these findings, ranging from heavy seeds to light seeds experiencing bursts of high accretion rate \citem{Schneider2023, Trinca2022, Volonteri2023,Pacucci2023,Bennett2024,Zhang23,Koudmani23}.
Yet, current datasets are unable to differentiate between these various scenarios.
Here we report the detection, from the JADES survey, of broad H$\alpha$ emission in a galaxy at z=6.68, which traces a black hole with mass of $\sim 4\times 10^8 M_\odot$ and accreting at a rate of only 0.02 times the Eddington limit.
The black hole to host galaxy stellar mass ratio is $\sim 0.4$, i.e. about 1,000 times above the local relation, while the system is closer to the local relations in terms of dynamical mass and velocity dispersion of the host galaxy. This object is most likely the tip of the iceberg of a much larger population of dormant black holes around the epoch of reionisation. Its properties are consistent with scenarios in which short bursts of super-Eddington accretion have resulted in black hole overgrowth and massive gas expulsion from the accretion disk; in between bursts, black holes spend most of their life in a dormant state.}

\keywords{Black Holes, Active Galactic Nuclei, Super-Eddington accretion, Galaxy Evolution}



\maketitle


The galaxy JADES GN+189.09144+62.22811 1001830  hereafter GN-1001830, located in the GOODS-N field,
was observed with JWST both with NIRCam and with the NIRSpec multi-object mode, both with the low resolution prism and medium resolution gratings as part of the JADES (JWST Advanced Extragalactic Survey), PID:1181. The NIRSpec spectra reveal multiple emission nebular lines (Fig. \ref{fig:full_prism}), which unambiguously show that the galaxy is at $z = 6.677\pm 0.004$.

\begin{figure}
    \centering
    \includegraphics[width=\textwidth]{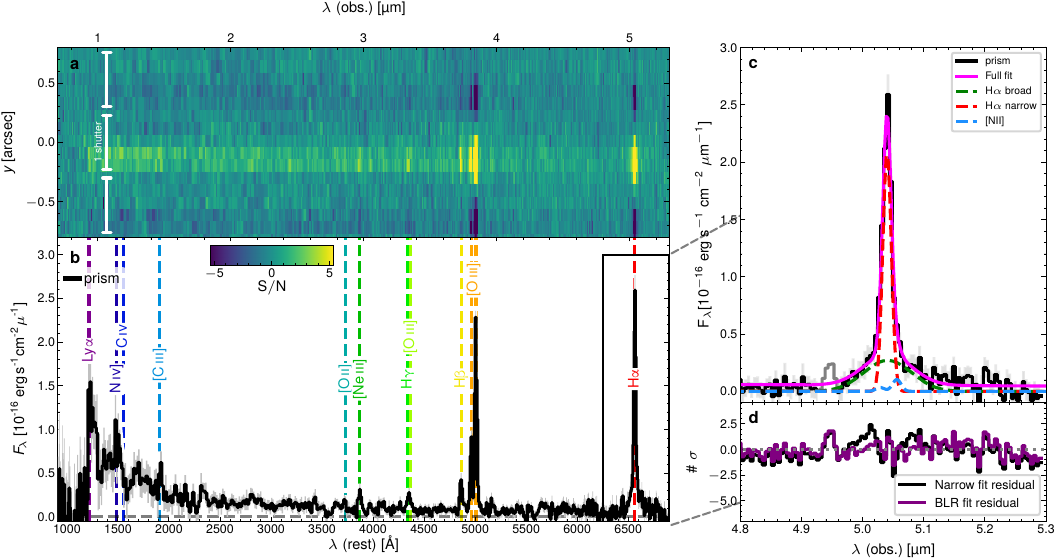}
    \caption{\textbf{Prism spectrum and \Has line of GN-1001830.} \textbf{a} shows the 2D prism spectrum, \textbf{b} -  the 1D prism (black line in the bottom panel) with marked emission lines. The spectrum around H$\alpha$ is shown in \textbf{c} illustrating the presence of a broad component. Lines shown are the observed spectrum (black solid line, with grey shading indicating $1\sigma$ uncertainties) along with the best fit line to the narrow (red dashed) and broad (green dashed) components. The [NII] doublet is shown in blue and it is only marginally detected at ~2$\sigma$. The magenta solid line shows the total fit. The greyed line portion at $\sim 4.95\ \mu$m of the spectrum shows the region, which was masked due to a possible artefact or \Has emission from a lower redshift interloper. \textbf{d} shows the fit residuals for a simple narrow \Has and \NIIs fit (black line) and the best-fit, containing a broad component (purple line). It can be seen that the narrow line only fit does not account for the broad wings of the line, leaving significant systematic residuals.}
    \label{fig:halpha_fit}
\end{figure}

The H$\alpha$ line is in the gap of the medium resolution grating spectrum and observed only in the prism spectrum (Fig.\ref{fig:halpha_fit}). However, the resolution at this wavelength
 is sufficient to reveal a clear broad component of this line. The broad component is fairly symmetric and not seen in [OIII]
(Fig.\ref{fig:OIII_Hbeta}). This suggests that the broad \Has line is not associated with outflows, leaving as the most plausible interpretation the Broad Line Region (BLR) of an accreting black hole (BH), i.e. an Active Galactic Nucleus (AGN). 

The broad component has a width  
of $5700_{-1100}^{+1700}$~km~s$^{-1}$ and a flux of $27.3^{+4.1}_{-4.0}\times10^{-19}$~erg~s$^{-1}$~cm$^{-2}$.
Assuming the local virial relations \citem{Reines2013,VolonteriBHmass}, and taking into account the effect of dust obscuration, we estimate a black hole mass of $\log{(M_{\rm BH}/M_{\odot})} = 8.61^{+0.38}_{-0.37}$ (see Methods).

Coupled with the bolometric luminosity, estimated from the broad component of H$\alpha$, 
but also consistently from the photometric fit of the nuclear component (see Methods),
we infer that the galaxy is accreting at 2.4\% of its Eddington limit, i.e.
$\lambda_{\rm Edd} \equiv L_{\rm bol}/L_{\rm Edd} = 0.024^{+0.011}_{-0.008}$, with an intrinsic scatter of 0.5~dex (see Methods for details).
   
The fact that the AGN is so underluminous allows constraining the properties of the host galaxy much better than in luminous quasars.  
We employed the \texttt{ForcePho} (B. Johnson, in prep.) tool to decompose the contribution of the nuclear region, hosting the unresolved AGN, from the host galaxy in NIRCam images (see Methods section). 
The morphology can be well fitted with a nuclear unresolved source and a compact ($R_e\sim 140~ {\rm pc}$) host galaxy with a disc-like profile (S\'ersic index $n\sim 1$).

The 8 bands photometry of the host galaxy was then fitted with the SED fitting codes  \texttt{BAGPIPES} \citem{Carnall2018} and \texttt{Prospector} \citem{Prospector} (see Methods). 
These fits are consistent and averaged together, yield a stellar mass $\log{(M_*/M_\odot)} = 8.92^{+0.30}_{-0.31}$ and instantaneous star formation rate $\rm SFR = 1.38^{+0.92}_{-0.45}$~M$_\odot$~yr$^{-1}$(within the last 10~Myr), which places our object a factor of 3 below the star-forming main sequence at its redshift.

\begin{figure}
    \centering
    \includegraphics[width=1.0\textwidth]{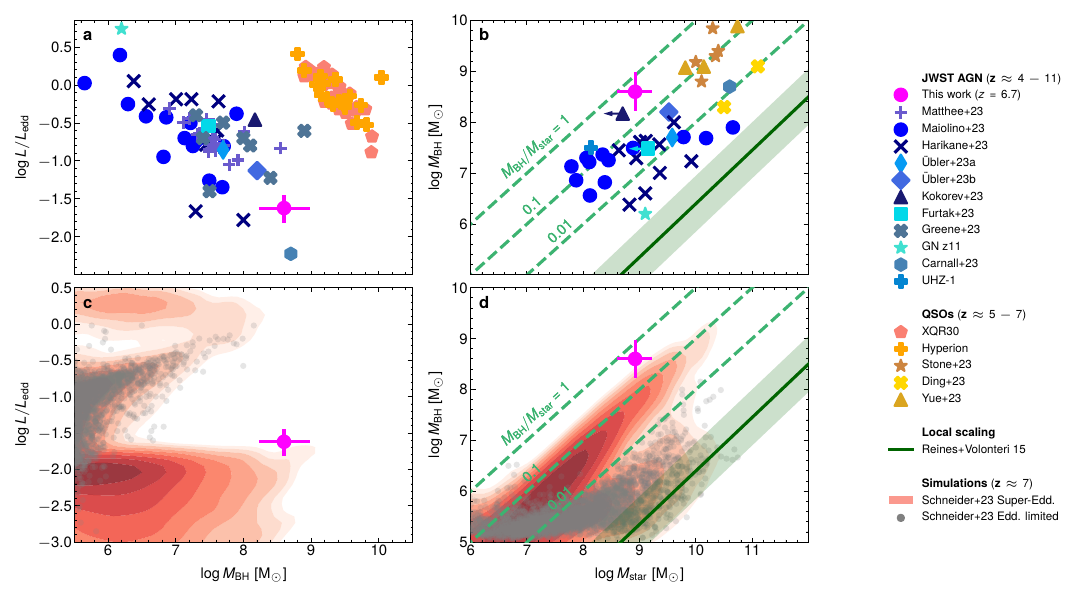}
    \caption{\textbf{Comparison of GN-1001830 with other high-z AGN and models in terms of accretion rate, black hole mass and stellar mass of the host galaxy.} {\textbf{Left panels (a, c):}} accretion rate relative to the Eddington limit, $\lambda_{\rm Edd}$, versus black hole mass, 
    $\log{M_{\rm BH}}$. {\textbf{Right panels (b, d):}} Black hole mass versus stellar mass of the host galaxy 
    $\log{M_{\rm star}}$. Green dashed lines indicate constant $M_{\rm BH}/M_{\rm star}$ ratios, while the solid green line represents the local relation from \protect\citem{VolonteriBHmass}, the shaded region shows the scatter.
    In all panels GN-1001830 is indicated with a magenta circle with errorbars.
    {\textbf{Top panels (a, b):}} Comparison with other JWST-discovered AGN at high redshift (blue symbols \protect\citem{Harikane_AGN, Maiolino_AGN, Matthee2023,Ubler2023, CArnall2023_AGN, Goulding2023_AGN, Kokorev2023_AGN,Ubler23a, Furtak2023_AGN}) and with the QSO population at similar redshifts (orange/yellow symbols \protect\citem{XQR30, Hyperion,Stone2023_QSO, Ding2023_QSO, Yue2023_QSO}). The observed negative correlation between $\lambda_{\rm Edd}$ and $M_{\rm BH}$ is likely reflective of Eddington luminosity's dependence on black hole mass and observational incompleteness and not a separate physical phenomenon. {\textbf{Bottom panels (c, d):}} 
    Comparison with the predictions (at z$\sim$7) from the semi-analytical models from \protect\citem{Schneider2023, Trinca2022} in the scenario of Eddington-limited accretion (grey points) and the scenario of light or heavy seeds that can experience super-Eddington accretion (red shaded contours).}
    \label{fig:comparisons}
\end{figure}

Fig.\ref{fig:comparisons} shows the location of GN-1001830 (large magenta circle) on the
$L/L_{\rm Edd}$ verus $M_{\rm BH}$ diagram (left panels) and on the $M_{\rm BH}$ versus $M_{\rm star}$ diagram (right panels).
On the top panels, our source is compared with 
other AGN discovered by previous JWST studies at similar redshifts (4$<$z$<$11, blue symbols) \citem{Harikane_AGN, Maiolino_AGN, Matthee2023, Ubler2023, CArnall2023_AGN, Goulding2023_AGN, Kokorev2023_AGN, Ubler23a, Furtak2023_AGN}, along with bright $z > 5$ QSOs observed with JWST (orange/yellow symbols) \citem{XQR30, Hyperion, Stone2023_QSO, Ding2023_QSO, Yue2023_QSO}.
The top-left panel (Fig.\ref{fig:comparisons}a) illustrates that 
our object is among the most massive black holes found by JWST, with a mass similar to that of luminous high redshift quasars, but it accretes at a rate lower by about 2 orders of magnitude. Therefore, our object is the dormant counterpart of luminous, high redshift quasars.

Additionally, the top-right diagram (Fig.\ref{fig:comparisons}b) indicates that our object is one of the most ``overmassive'' BHs found by JWST, in the sense that the BH mass approaches 50\% of the host's stellar mass, i.e. about 1,000 times above the local relation between BH and host galaxy stellar mass. 

\begin{figure}
    \centering
    \includegraphics[width=\textwidth]{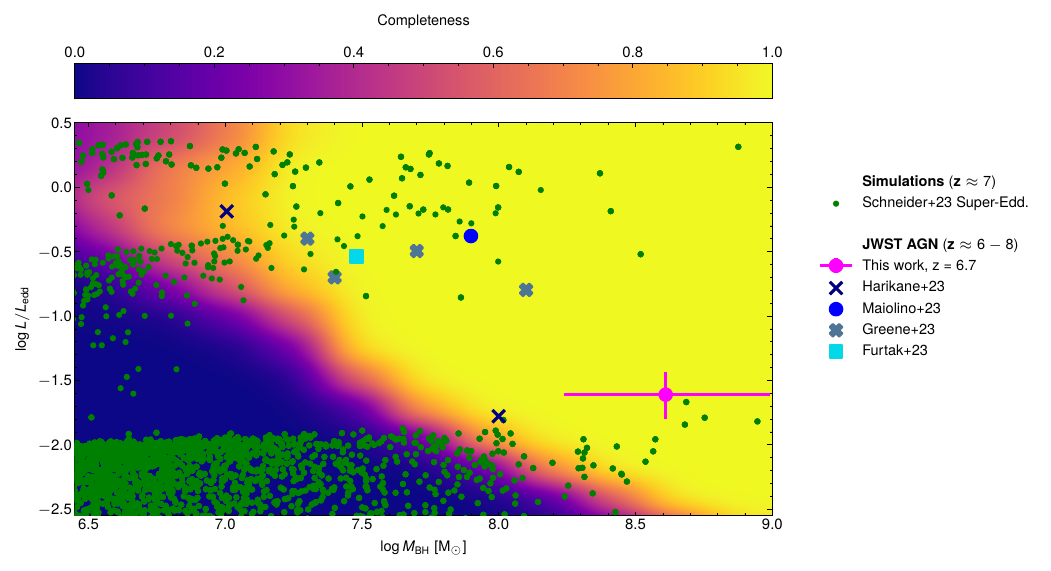}
    \caption{\textbf{Completeness simulation results on the Eddington ratio versus black hole mass plane.} Points colored in blue hues show the previously discovered JWST sources at 6$<$z$<$8, as in Fig.\ref{fig:comparisons}. The dark green points show the simulated AGN (at z$\sim$7) in the scenario of super-Eddington bursts. GN-1001830 is indicated with magenta circle with errorbars. The color shading indicates the completeness of the JADES spectroscopic survey in detecting black holes with a given mass and accreting at a given rate relative to Eddington. It can be readily seen that most of the low accretion rate AGN predicted by super-Eddington bursts lie in the sub-50\% completeness region and that GN-1001830 overlaps them at the edge of the high completeness region.}
    \label{fig:completeness}
\end{figure}

The JWST finding of several overmassive black holes at high redshift \citem{Ubler2023, Harikane_AGN, Maiolino_AGN, Bogdan2022,Maiolino24_GN-z11} has been interpreted by some works \citem{TrinityIV,Li2024} as the result of a large scatter
of the BH-stellar mass relation
combined with selection effects, in the sense that more massive black holes tend to be preferentially selected, as they can reach higher luminosities.
Our discovery of such highly overmassive black hole associated with a low luminosity AGN, due to its low Eddington ratio, is incompatible with the selection effect scenarios as our data is deep enough to be less sensitive to selection effects (Fig.\ref{fig:completeness}). 
This is discussed more extensively in the Methods.

Some previous studies have found that early black holes are overmassive only relative to the stellar mass, but when compared with the velocity dispersion and dynamical mass of the host galaxy they are more aligned with the local relation (\citem{Maiolino_AGN,Gravity24}). 
As detailed in the Methods, based  on the profile of the \OIIIs doublet, we find that this is indeed the case also for GN-1001830: in contrast to its strong offset on the $M_{BH}-M_{star}$, this galaxy is closer to the local $M_{BH}-\sigma$ and $M_{BH}-M_{dyn}$ relations.
This may indicate that the baryonic mass of the host galaxy is already in place, but that star formation is lagging behind, possibly due to feedback generated by black hole accretion. 

The presence of overmassive black holes in the early Universe has been explained by a variety of models and cosmological simulations. These predict that either black holes are born from relatively massive seeds (often called `heavy seeds', such as direct collapse black hole, originating from clouds of pristine gas) accreting below the Eddington rate, or from short phases of super-Eddington accretion (possibly driven by galaxy mergers) either on light (stellar remnants) or heavy seeds  (\citem{Schneider2023, Trinca2022, Bennett2024, Volonteri2023, Pacucci2023, Trinca2023, Koudmani23, Dayal2024, Zhang23, Lupi23, Pezzulli2016}).
The bottom panels of Fig.\ref{fig:comparisons}  show the comparison of  GN-1001830 with the CAT semi-analytical models from \citem{Schneider2023, Trinca2022}, who envisage both scenarios, in a snapshot at $z=7$. The Eddington-limited, heavy-seeds scenario (grey small symbols) fails to reproduce the properties of GN-1001830. Indeed, in order to reproduce the large black hole masses observed at high redshift without exceeding the Eddington limit, this scenario requires black holes to be accreting close to the Eddington limit for most of the time.  Therefore, the high mass black holes, predicted by this scenario, are not found at the significantly sub-Eddington rates as observed in our object. Additionally, this scenario can either reproduce a very overmassive nature of black holes ($M_{\rm BH}/M_{\rm star}\sim 0.1$) only for low mass galaxies ($M_{\rm star}\sim 10^6-10^7~M_\odot$), or more moderately overmassive black holes ($M_{\rm BH}/M_{\rm star}\sim 0.01$) in more massive galaxies, hence is not capable of reproducing the $M_{\rm BH}/M_{\rm star}=0.43$ observed in GN-1001830 with $M_{\rm star}=2\times 10^9~M_\odot$. On the contrary, the models show that, even starting from light seeds, allowing super-Eddington accretion bursts (red shaded contours)  can  reproduce the observed properties of our object. It may sound counter-intuitive that super-Eddington scenarios can better reproduce  the relatively quiescent AGN in GN-1001830. The fact is that super-Eddington accretion phases allow the black hole to grow rapidly in short (1-4 Myr) bursts, while the resulting strong feedback makes the black hole lack gas to accrete significantly for long periods. This way, black holes can reach high masses while staying dormant for long periods, increasing the probability of seeing them in a low luminosity (dormant) state.

We note that the same result is found also when comparing with fully self-consistent cosmological simulations of galaxy formation, such as  FABLE, as discussed in the Methods.

In addition to the super-Eddington scenario described above, models invoking radiatively inefficient accretion onto low spin black holes \citem{Inayoshi2024} could also help in explaining our finding. However, the detailed treatment of this scenario is beyond the scope of our work.

It is tempting to speculate that our result favors light seed models. However, the same result would also hold if the models had started with heavy seeds. The key feature that allows the properties of GN-1001830 to be matched is the fact that accretion goes through super-Eddington phases, regardless of the seeding mechanism.

Finally, we argue that dormant, overmassive black holes in galaxies with low SFR, such as GN-1001830, are probably quite common in the early Universe. Indeed, finding one of them out of 35 spectroscopically targeted galaxies at $z > 6$ in the GOODS-N field, in a single tier of the JADES survey, is remarkable, as the JADES selection function at $z > 6$ disfavours the selection of high-z galaxies with low star formation rates \citem{Bunker2023_nirspec}. Additionally, the very low BH accretion rate makes the intensity of the broad lines very weak and much more difficult to detect relative to all other AGN found at high-z. 

 The fact that, out of the three Type 1 AGN at $z > 6$ currently found by JADES, one is a dormant black hole in a relatively quiescent galaxy, despite all the selection effects against this class of objects, indicates that they must be much more numerous and much more common than actively accreting AGN in star forming galaxies.
 We have performed a completeness simulation to infer the capability of the JADES survey in detecting black holes with a given mass and accretion rate, at the same redshift as GN-1001830 (see Methods for details). The results are shown in Fig.\ref{fig:completeness}, where light background colors indicate higher levels of completeness. As expected, at a given black hole mass, black holes accreting more vigorously are easier to detect. The comparison with the same simulations as in Fig.\ref{fig:comparisons} (green points, from \citem{Schneider2023, Trinca2022}) reveals that GN-1001830 overlaps with the high-mass tail of dormant black holes in the region where a few of these become detectable in the JADES survey. However, this is just the tip of the iceberg, as the majority of these dormant black holes are expected to be undetected. Specifically, only 0.1\% of the simulated BHs from \citem{Schneider2023, Trinca2022} with masses lower than $10^8$~M$_{\odot}$ and Eddington ratios below 0.03 are detectable in the JADES survey. This fraction becomes $\sim$50\% in the higher BH mass range tail probed by GN-1001830 ($10^8 < M_{\rm BH} < 10^9$~M$_{\odot}$). It should also be noted that the presence of GN-1001830 in the $\sim$100 arcmin$^{2}$ of GOODS-N field implies a number density of $\sim 10^{-5.2}$~Mpc$^{-3}$, which is consistent within a factor of two with the simulations prediction of $\sim 10^{-4.9}$~Mpc$^{-3}$ \citem{Trinca2023}, especially given that we have not spectroscopically targeted all possible AGN in the fieds. The plot also confirms that the several black holes found by JWST accreting close to the Eddington rate (blue symbols) are preferentially selected in this phase only because they are much more luminous and easier to detect. Some of the AGN observed at high redshift are seen accreting at super-Eddington \citem{Maiolino_AGN,Gravity24, Fujimoto2022} rates, confirming the existence of (short) super-Eddington phases.
GN-1001830 is detected, despite being dormant, because it is just above the detectability threshold; yet, our result suggests that the majority of black holes at high redshift are dormant and apparently rare only because they are much more difficult to detect. 


\bmhead{Acknowledgments}
The authors would like to thank M. Volonteri and A. Fabian for 
helpful comments and useful advice. S.A. acknowledges support from Grant PID2021-127718NB-I00 funded by the Spanish Ministry of Science and Innovation/State Agency of Research (MICIN/AEI/ 10.13039/501100011033). W.B. acknowledges support by the Science and Technology Facilities Council (STFC), ERC Advanced Grant 695671 "QUENCH". AJB, AJC, JC, IEBW, AS and GCJ acknowledge funding from the "FirstGalaxies" Advanced Grant from the European Research Council (ERC) under the European Union’s Horizon 2020 research and innovation programme (Grant agreement No. 789056). S.C acknowledges support by European Union’s HE ERC Starting Grant No. 101040227 - WINGS. RM acknowledges support by the Science and Technology Facilities Council (STFC), by the ERC through Advanced Grant 695671 "QUENCH", and by the UKRI Frontier Research grant RISEandFALL. RM also acknowledges funding from a research professorship from the Royal Society. J.S.B. acknowledges support from the Simons Collaboration on `Learning the Universe'. D.S. acknowledges support by the Science and Technology Facilities Council (STFC). M.P. acknowledges support from the research project PID2021-127718NB-I00 of the Spanish Ministry of Science and Innovation/State Agency of Research (MICIN/AEI/ 10.13039/501100011033), and the Programa Atracci\`on de Talento de la Comunidad de Madrid via grant 2018-T2/TIC-11715. BER acknowledges support from the NIRCam Science Team contract to the University of Arizona, NAS5-02015, and JWST Program 3215. The research of CCW is supported by NOIRLab, which is managed by the Association of Universities for Research in Astronomy (AURA) under a cooperative agreement with the National Science Foundation. The FABLE simulations were performed on the DiRAC Darwin Supercomputer hosted by the University of Cambridge High Performance Computing Service (http://www.hpc.cam.ac.uk/), provided by
Dell Inc. using Strategic Research Infrastructure Funding from the Higher Education Funding Council for England and funding from the Science and Technology Facilities Council; the COSMA Data Centric system at Durham University, operated by the Institute for Computational Cosmology on behalf of the STFC DiRAC HPC Facility.This equipment was funded by a BIS National E-infrastructure capital grant ST/K00042X/1, STFC capital grant ST/K00087X/1, DiRAC Operations grant ST/K003267/1 and Durham University. S.K. acknowledges support from St Catharine's College through a Junior Research Fellowship. RS acknowledges support from the PRIN 2022 MUR project 2022CB3PJ3 - First Light And Galaxy aSsembly (FLAGS) funded by the European Union – Next Generation EU. AT and RV acknowledge support from the PRIN 2022 MUR project 2022935STW and 2023 INAF Theory Grant "Theoretical models for Black Holes Archaeology". M.A.B. acknowledges support from a UKRI Stephen Hawking Fellowship (EP/X04257X/1) as well as from the Science and Technology Facilities Council (STFC).

\bmhead{Author contributions}
I.J., R.M., and J.S. contributed to the analysis, and initial interpretation of the spectroscopic data. I.J. performed the completeness simulations. All authors contributed to the interpretation of results. S.A., S. Carniani, M.C., J.W. and M.P. contributed to the NIRSpec data reduction and to the development of the NIRSpec pipeline. S.A. contributed to the design and optimization of the MSA configurations. S.T. and W. M. B. contributed to the analysis and interpretation of the NIRCam imaging data. R.S., A.T. and R.V. contributed with the raw and advanced data products from the CAT simulations. J.S.B, M.A.B, B.J., S.K. and D.S. contributed with the advanced data products from their FABLE simulations. C.D. performed the variability analysis of the source. A.G., J.S. and F.D. contributed to the development of tools for the spectroscopic data analysis and visualization. B.R. contributed to the JADES data reduction.

\bmhead{Data Availability}
The reduced data used to make the figures together with the unprocessed data has been made available on the STScI archive as part of JADES Data Release 3 (https://archive.stsci.edu/hlsp/jades).

\bmhead{Author Information}
The authors declare that they have no competing financial interests. Correspondence and requests for materials should be addressed to I.J. (email: ij284@cam.ac.uk).

\bibliographym{sn-bibliography}

\newpage

\section{Methods}
\subsection{Data description and reduction}

All data used herein has been obtained from the JADES survey, the full description of which is available in \citesup{JADES_desc}. The spectroscopic survey consists of several tiers, characterized by their depth, 'Medium' or 'Deep', photometry from which targets were selected, HST or JWST, and the observed field - GOODS-S or GOODS-N. Here we make use of data from the Medium/HST in GOODS-N tier, which consists of a single Near Infrared Spectrograph (NIRSpec) spectrum in prism and R1000 gratings. We also make use of the accompanying Near Infrared Camera (NIRCam) wide-band imaging data. 

\subsubsection{NIRSpec}
The full description of the NIRSpec data used is available in \citem{Maiolino_AGN}. A brief summary is  provided here for completeness. The observations in the Medium/HST tier in GOODS-N consisted of three medium resolution gratings (G140M/F070LP, G235M/F170LP, G395M/F290LP) and low resolution prism. The exposure time was 1.7 hours per source in prism and 0.8 hours per source in the medium gratings. The data were processed according to procedures laid out in \citesup{Bunker2023} and other similar JADES papers, such as \citesup{Carniani2023}. A full description of the data reduction procedure will be presented in Carniani et al. in prep, here we note that the spectral data was reduced using the pipeline developed by NIRSpec GTO team and the ESA NIRSpec Science
Operations Team. As the primary interest of this study was the properties of the central, unresolved region, containing the AGN, we use the 1D spectra extracted from the central 3 pixels, corresponding to 0.3", of each 2D spectra. Path-loss corrections were calculated for each observation, taking into account the intra-shutter position, assuming a point-source geometry and a 5-pixels extraction box. Due the compact nature of the object, to maximise the S/N, we use 3 pixel extractions; even though a 3-pixel box is not the extraction box we optimised the path-loss corrections for, we compared directly the two spectra and found no systematic difference within the uncertainties.

\begin{figure}
    \centering
    \includegraphics[width=\textwidth]{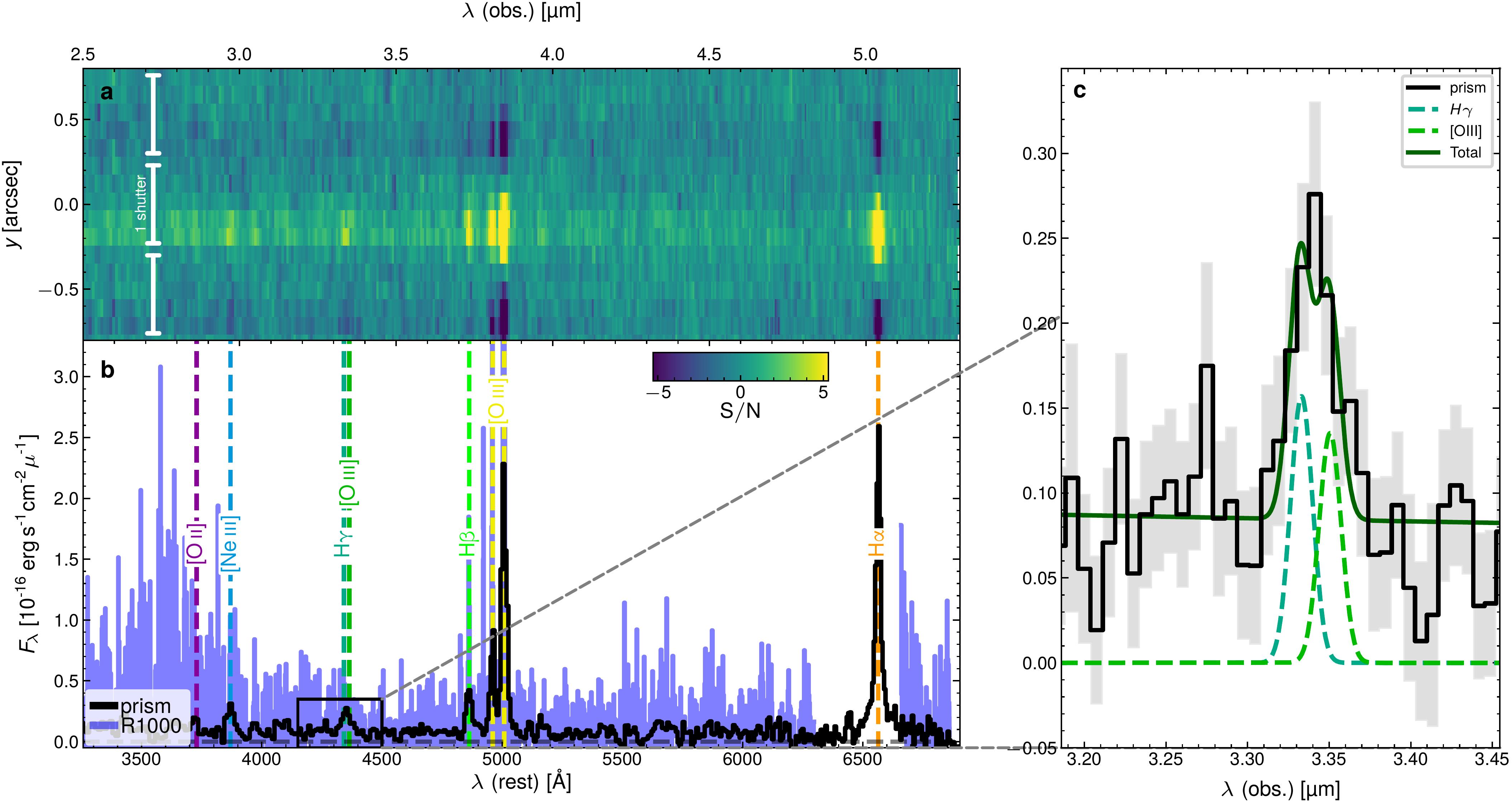}
    \caption{\textbf{Prism spectrum of GN-1001830.} The top panel shows the 2D spectrum with the y axis representing the shutter pitch and yellower portions showing more positive flux. The bottom panel shows the extracted 1D prism spectrum with emission line locations indicated by colored vertical lines. The (noisier) R1000 spectrum is shown in blue, the wavelength range is narrowed with respect to Fig.\ref{fig:halpha_fit} to leave out the noisiest parts of R1000. The panel to the right shows a zoomed-in view on the blended \Hg\ and \OIII$\lambda$4363 feature along with its decomposition into two Gaussian profiles.}
    \label{fig:full_prism}
\end{figure}

The full obtained prism spectrum is shown in Fig.\ref{fig:full_prism}.

\subsubsection{NIRCam}

\begin{figure}
    \centering
    \includegraphics[width=0.6\linewidth]{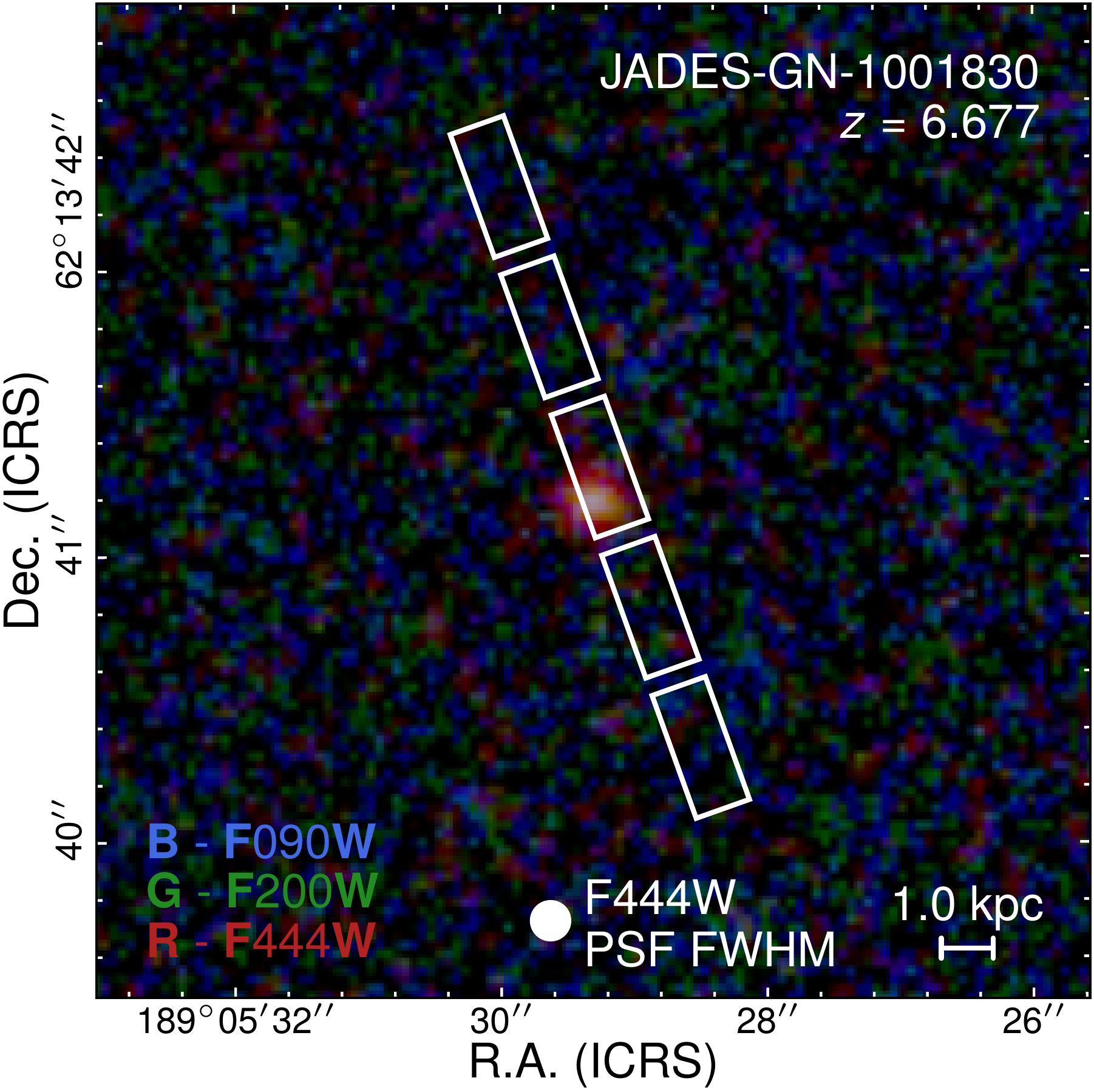}
    \caption{\textbf{Image of GN-1001830.} A red-green-blue (RGB) image of the AGN and galaxy in the F444W, F277W and F115W bands. A 1 kpc physical scale bar is overplotted alongside the FWHM of the PSF in the F444W band. The position of the NIRSpec slit is also overplotted.}
    \label{fig:rgb}
\end{figure}

The imaging data consisted of 7 wide (F090W, F115W, F150W, F200W, F277W, F356W and F444W) and 1 medium (F410M) filter bands of the NIRCam instrument in the GOODS-N field. We also use imaging in the F182M and F210M medium bands.
The photometric data reduction procedure is presented in \citesup{Robertson2023, Tacchella2023} and \citesup{Baker2023} with a full description to be made available in Tacchella et al. in prep. In summary, we use v1.9.2 of the JWST calibration pipeline \citesup{bushouse_2022_7229890} together with the CRDS pipeline mapping context 1039. Stages 1 and 2 of the pipeline were run with our own sky-flat provided for the flat-fielding, otherwise keeping to the default parameters. After Stage 2, custom procedures were performed to account for $1/f$ noise and subtract scattered light artifacts, "wisps", along with the large scale background. Astrometric alignment was performed using a custom version of JWST TweakReg, with corrections derived from HST F814W and F160W mosaics along with GAIA Early Data Release 3 astrometry. The images of individual exposures were then stacked in Stage 3 of the pipeline with the final pixel scale being 0.03" per pixel. An RGB image of our galaxy is shown in Fig. \ref{fig:rgb}

\subsection{Spectral fitting and further spectral analysis}
\label{sub:spec_fit}
In order to identify the broad component in the H$\alpha$ line, we use a Bayesian method to model it with two components models - one containing only narrow emission in the H$\alpha$ line, \NII$\lambda$$\lambda$6548,6583 and [SII]$\lambda$$\lambda$6716,6731 doublets, the broad-line model included a broad component in the H$\alpha$ line. Narrow-line widths were constrained to be the same for every line, the ratio of the [NII] doublet fluxes was fixed to 3 (as from their Einstein coefficients ratios), the [SII] doublet fluxes remained independent but constrained to be within the flux ratio 6716/6731 range expected in the low and high density regimes (0.45 - 1.45, \citesup{Wang2004}). The priors on the peak widths were uniform with the fitted FWHM ranging between 700 and 1,500~km~s$^{-1}$ for the narrow, and between 1,500 and 11,500~km~s$^{-1}$ for the broad component, with the lower bound set by instrument resolution The posterior is estimated with a Markov-Chain Monte-Carlo integrator, \citesup{emcee}.  Redshifts for the narrow peaks and the BLR were fit independently with priors being set to narrow Gaussians centered on the overall redshift obtained through visual inspection and widths inferred from the pixel scale in redshift space. Line peak heights utilized log-uniform priors.

\begin{figure}
    \centering
    \includegraphics[width=\linewidth]{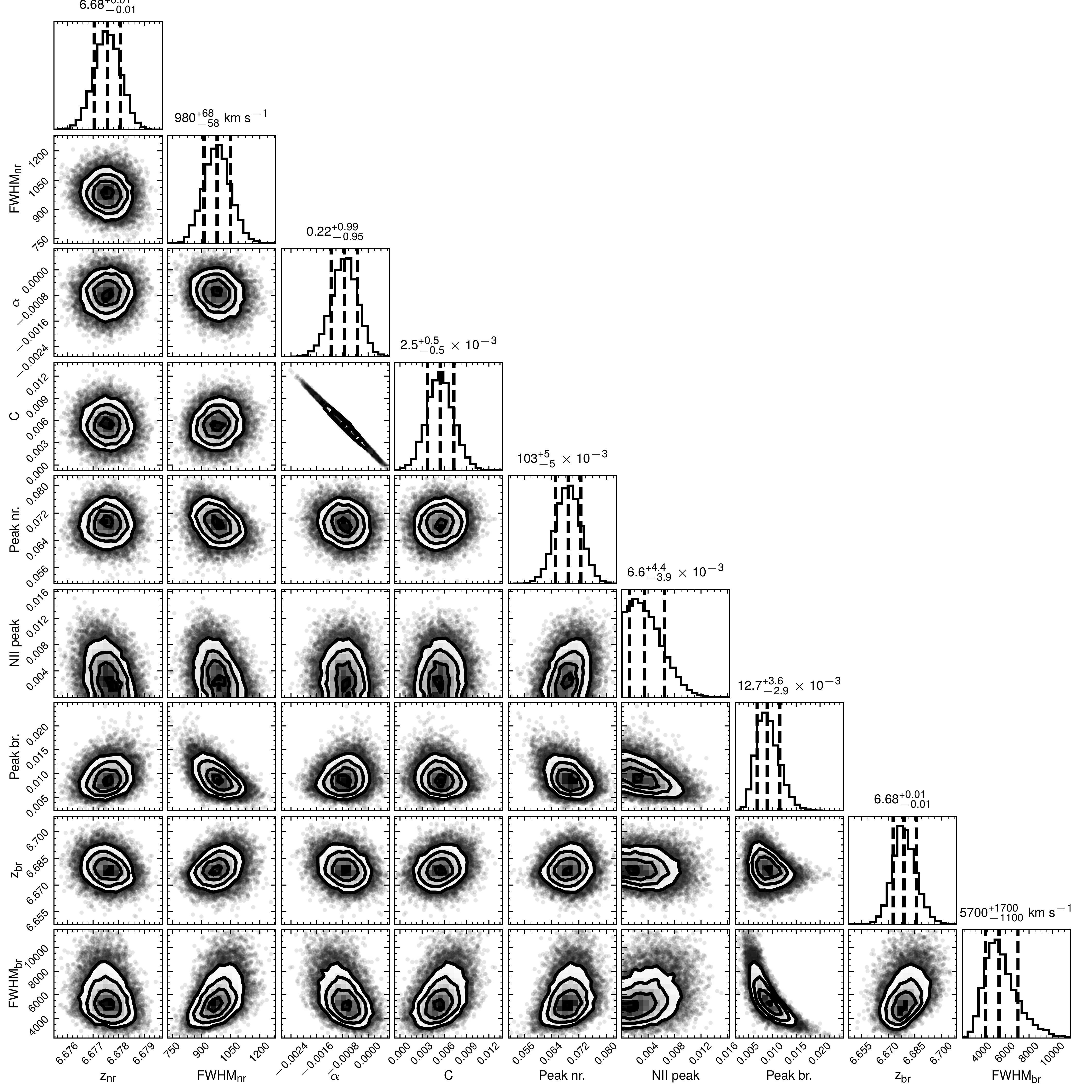}
    \caption{\textbf{A corner plot summarizing the MCMC fit results of the H$\alpha$ profile.} Here $\alpha$ and C are the continuum slope and normalization respectively, "Peak nr." and "Peak br." show the peak heights of the narrow and broad \Has components respectively. The FWHM values given are not corrected for instrumental broadening. The total spectrum was normalized to unity before fitting, hence peak heights are unitless here. The \SIIs doublet parameters are not shown as it is undetected.}
    \label{fig:corner}
\end{figure}

Performance of the two models was quantified using the Bayesian information criterion (BIC), defined as
\begin{equation}
     BIC = \chi^2 + k\ln{n,}
\end{equation}
where $k$ is the number of free parameters and $n$, the number of data points fitted. Following the criteria in \citem{Maiolino_AGN}, we require  $\Delta BIC = BIC_{\rm Narrow} - BIC_{\rm Narrow+Broad}$ to be above 5 for a robust detection. In addition, we require the fitted broad component to have a significance of at least 5$\sigma$. The significance of the broad component of GN-1001830 was found to be 9$\sigma$ and $\Delta BIC$ came out to be 31. The full summary of the fit is shown in a corner plot in Fig.\ref{fig:corner}. As can be seen there, the data is quite constraining on the different components with no significant degeneracies between them.

\begin{figure}
    \centering
    \includegraphics[width=\textwidth]{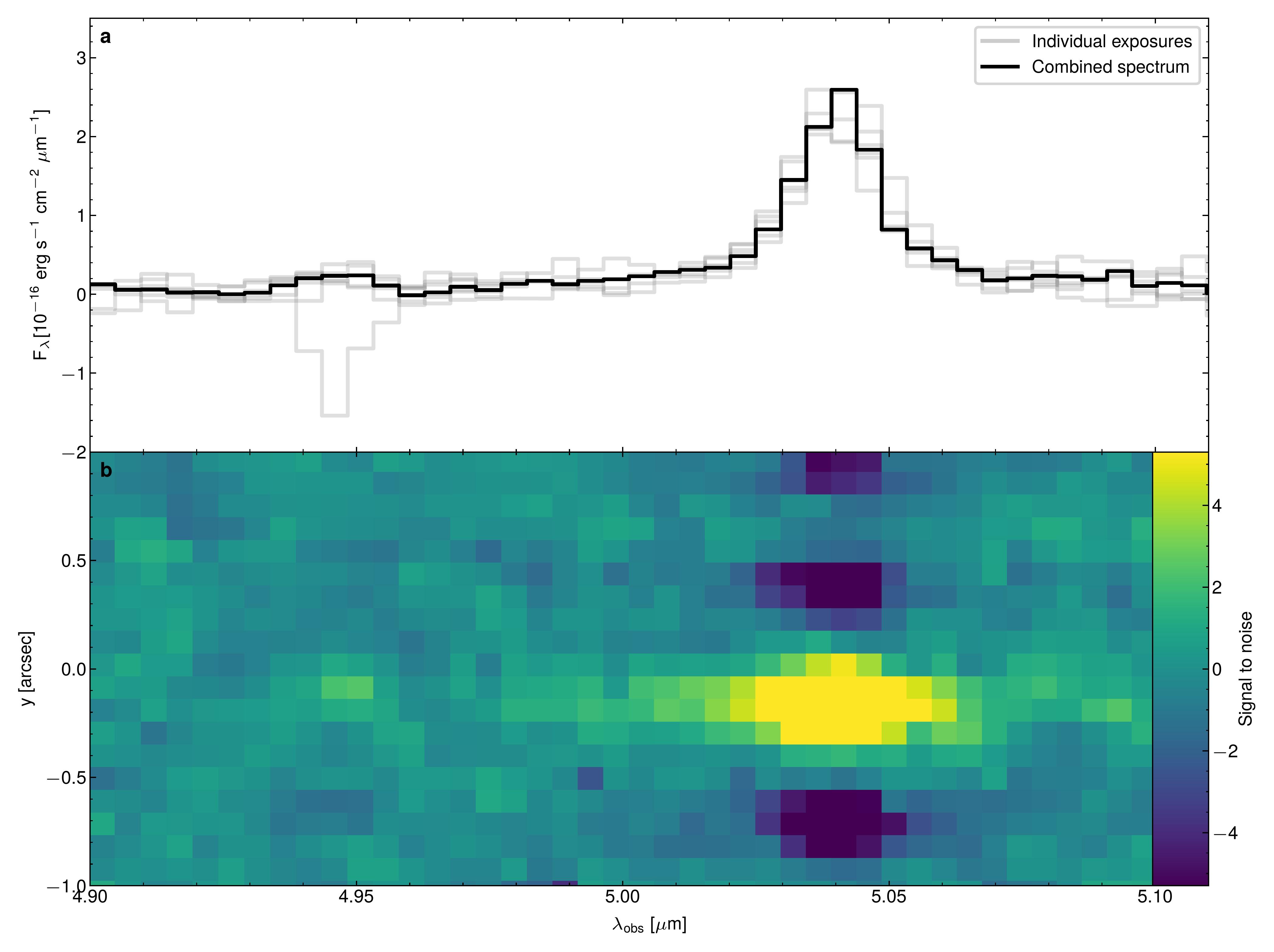}
    \caption{Combined spectrum around H$\alpha$ compared with the individual exposures. Grey lines in the top panel show the individual exposures, with the stacked spectrum shown in black. The bottom panel shows the 2D spectrum zoomed in on the same region. It can be seen that there is an outlier in the location of the artefact at $\lambda \approx 4.94\ \mu$m. The 2D cutout also shows the slight spatial offset of the feature.}
    \label{fig:artefact}
\end{figure}

We note that the spectrum shown in Fig.\ref{fig:halpha_fit} appears to have a feature at $\lambda \approx 4.94\ \mu$m, the origin of which is currently unclear. In the 2D spectrum, this feature is offset by 1-2 pixels, suggesting that it might be  \Has line of a foreground galaxy at z~$\approx 6.53$. However, one of the four exposures has an outlier at this location Fig.\ref{fig:artefact}; although this has been masked, the feature might be a residual artefact. For these reasons, we choose to mask this feature in the final fit, although its inclusion would not significantly affect the results. We also note that the \Hbs line in Fig.\ref{fig:full_prism} appears to contain a similar artefact, which manifests as an apparent broad wing. However, the inferred $\sim$2$\sigma$ significance of the feature leads us to conclude that its origin is noise.

\begin{figure}
    \centering
    \includegraphics[width=0.7\textwidth]{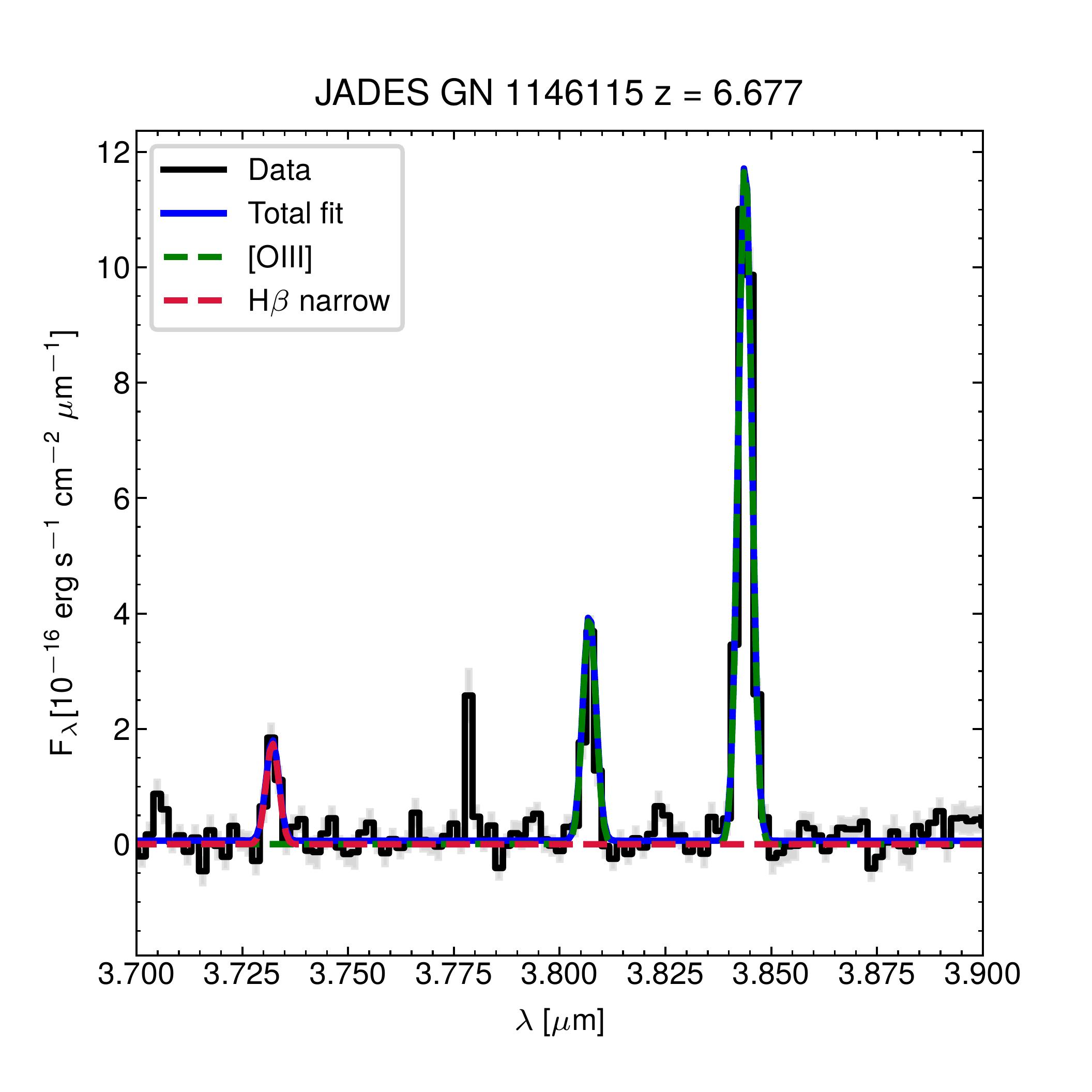}
    \caption{Grating spectrum zoomed around H$\beta$ and [OIII] along with the best fit model. It can be seen that the data is well explained by single component fits to each line, indicating no significant outflows. The spike at 3.775 $\mu$m is likely a noise feature that survived sigma clipping}
    \label{fig:OIII_Hbeta}
\end{figure}

To check for evidence of outflows, we fit the $H{\beta}$ line together with the \OIII \ doublet in the medium resolution data. For this purpose we fit these lines first with single components constrained to have the same width, with the ratio of [OIII] doublet peaks fixed at 3, then introduce a broader outflow component into each line, and finally we fit a broad component to $H{\beta}$. We find that single narrow component fits are preferred for each line, showing no evidence for outflows or a broad component in $H{\beta}$, as can be seen in Fig.\ref{fig:OIII_Hbeta}. The measured FWHM of the narrow lines in R1000 was 225$_{-11}^{+11}$~km~s$^{-1}$, when corrected for instrumental broadening, using the point source line spread function (LSF) models from \citesup{AnnaLSF}. This broadening is 180~km~s$^{-1}$ in the wavelength range considered.

\begin{figure}
    \centering
    \includegraphics[width=0.8\linewidth]{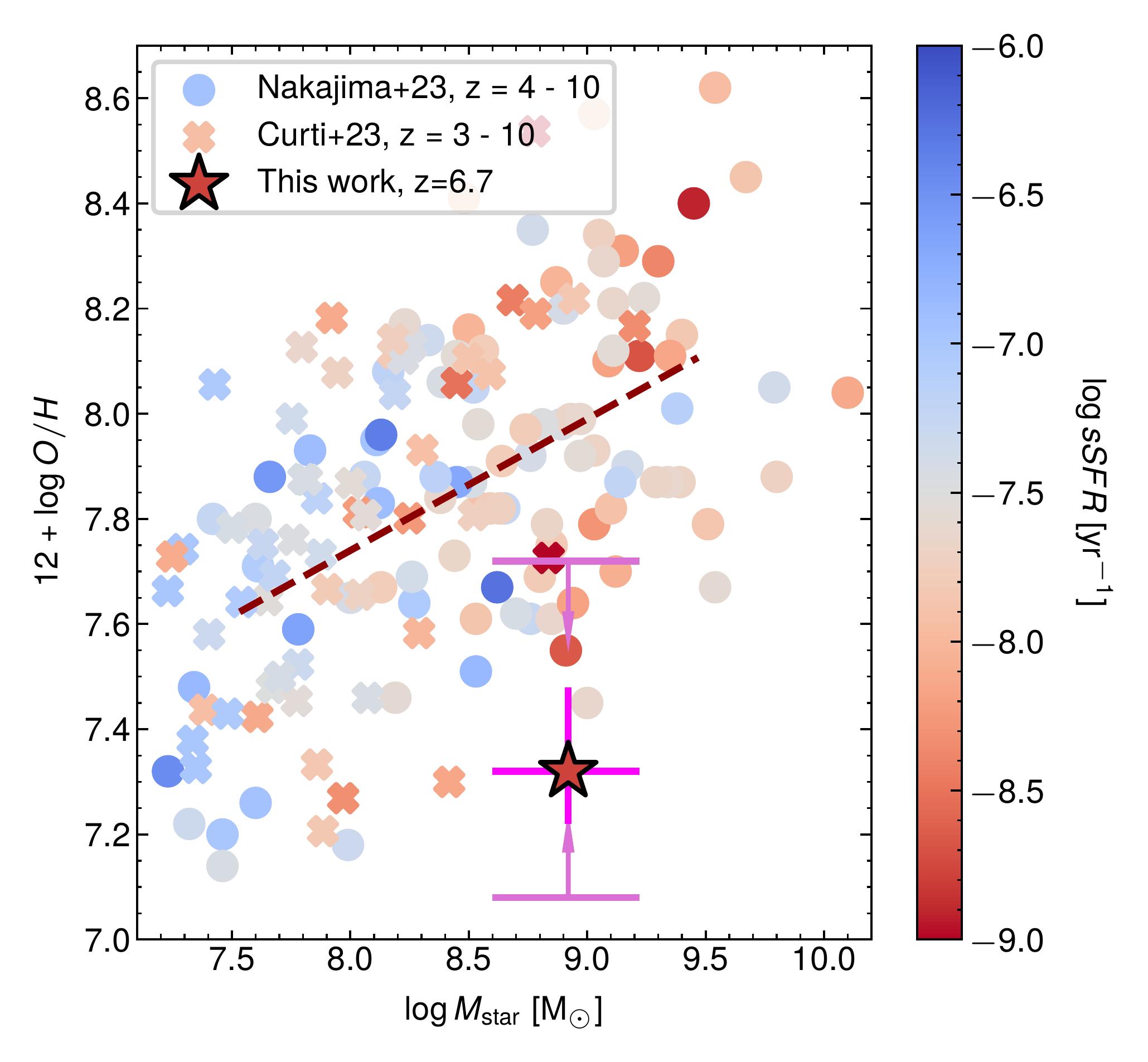}
    \caption{\textbf{Mass-metallicity relation} Location of GN-1001830 (large star with magenta errorbars) relative to the mass-metallicity relation at $z = 4-10$ obtained by \protect\citesup{Nakajima_metallicity} and \protect\citesup{Curti2023} (circles and crosses) whose best fit is shown with the brown dashed line. The lighter colored  magenta bars show the upper and lower limits for the metallicity of GN-1001830. All markers are colored according to their specific star formation rate, $\log(sSFR)$.}
    \label{fig:mass_metal}
\end{figure}

We also utilize previous Hubble Space Telescope (HST) imaging of the source carried out in 2018 to check its variability and thus the presence of supernovae that could potentially produce broad \Has emission while in their nebular phase. Subtracting HST and JWST images taken in equivalent filters shows that the flux of the source varied by no more than 5\% over the 4 year period in the observed frame, corresponding to $<$5\% variability over a 6 month period in the object's frame. This conclusively shows that our observed broad component can not be the product of a recent supernova explosion in the galaxy.
We also rule out the possibility that the broad-line component may be attributed to  the effect of multiple SNe given the low  star formation rate (SFR) of the galaxy.

In addition, we leverage the higher depth of the prism to fit weaker emission lines, in particular \Hg, \OIII$\lambda$4363, \NeIII$\lambda$3869 and \OII$\lambda$3727, as the \OIIIs auroral line may be used to constrain metallicity while the remaining lines are candidate diagnostics for Type 2 AGN. Each line has been fit with a single Gaussian profile together with \Hbs and the \OIIIs doublet with redshift and FWHM fixed by the latter. We note that, although the \OII$\lambda$3727 line is part of a doublet, said doublet is completely blended in the prism spectrum and the overall detection significance ends up being marginal (Table \ref{tab:lines}). All narrow emission lines fitted are summarized in Table \ref{tab:lines}. It should be noted that we carry out our fitting using spectra extracted from the central 3 pixels of the source in order to enhance the S/N of lines as the region of emission is compact due to the AGN nature of the source. However, the flux corrections applied by the reduction pipeline are geared towards the 5 pixel extracted spectra, we thus redo our fits using the 5 pixel spectra and find that the derived line fluxes differ by less than 1$\sigma$.

\begin{table}
    \centering
    \begin{tabular}{cccc}
    \toprule
       Line  & Flux ($\times 10^{-19}$ erg s$^{-1}$ cm$^{-2}$) & S/N & Disperser \\
    \hline
        \SII$\lambda$6731& $< 1.5$ & 1.38$\sigma$ & prism \\
       \NII$\lambda$6585  & $2.84_{-1.13}^{+1.06}$ & 2.0$\sigma$ & prism \\
       \Ha & $36.2_{-2.2}^{+2.3}$ & 25$\sigma$ & prism \\
         \OIII$\lambda$5007 & $46.7_{-1.4}^{+1.3}$  & 159$\sigma$ & R1000 \\
         \OIII$\lambda$4959 & $15.4_{-0.46}^{+0.43}$  & 65$\sigma$ & R1000 \\
         \Hb & $6.66_{-0.82}^{+0.81}$ & 29$\sigma$ & R1000 \\
         \OIII$\lambda$4363 & $2.39_{-0.65}^{+0.65}$ & 3.60$\sigma$ & prism \\
         \Hg & $2.77_{-0.70}^{+0.68}$ & 3.96$\sigma$ & prism \\
          \NeIII$\lambda$3869& $3.87_{-0.91}^{+0.94}$ & 3.63$\sigma$ & prism \\
          \OII$\lambda$3727 & $1.78_{-0.67}^{+0.66}$ & 2.64$\sigma$ & prism \\
    \bottomrule
    \end{tabular}
    \caption{\textbf{Summary of all narrow lines measured in GN-1001830.} Column one contains the names of each line, the second column - measured flux along with uncertainties, the final two columns contain signal to noise ratios and disperser in which each line was measured. For lines with S/N $<$ 3$\sigma$ the measured fluxes were treated as upper limits, for $<$ 2$\sigma$ detections, the 2$\sigma$ upper limit is quoted instead.}
    \label{tab:lines}
\end{table}

Our tentative (3.6$\sigma$) detection of  \OIII$\lambda$4363, together with the \OIII$\lambda$$\lambda$5007,4959 doublet can be used to estimate electron temperature based on their ratio. The electron density can not be reliably estimated from the spectrum due to lack of an \SII$\lambda$$\lambda$6716,6731 detection, however we note that the inferred electron temperature is relatively insensitive to density between 100 and 10,000~cm$^{-3}$, typical of narrow line regions of AGN \citesup{Zhang2024}. We thus assume electron density of order of 1000~cm$^{-3}$ and calculate the electron temperature to be $T_e \approx 25000^{+3200}_{-4800}$~K. Assuming the main ionization mechanism to be AGN activity we follow the methods from \citesup{Dors2020} to derive the contributions to metallicity from different ionic species of oxygen - $12 + \log\left(O^{++}/H\right) = 7.28_{-0.1}^{+0.16}$ and $12 + \log\left(O^{+}/H\right) < 6.13$, the latter value presenting an upper limit due to the low detection significance of \OII$\lambda$3727. The contribution of higher ionization oxygen species is likely negligible due to lack of \OIV\ and \HeII\ line detections. \HeII\ lines have similar ionization potential to \OIV\, but helium is much more abundant, thus their lack of detection implies that the radiation is not sufficiently hard to produce significant amounts of highly ionized oxygen. The final oxygen abundance ratio estimate is thus $12 + \log{\left(O/H\right)} = 7.32_{-0.10}^{+0.16}$, corresponding to $Z \approx 0.04Z_{\odot}$. However, the \OIII$\lambda$4363 line is heavily blended with \Hg\ in prism, therefore the simple two-Gaussian fit may underestimate the relevant uncertainties. As a conservative estimate we obtain a lower limit on the \OIII$\lambda$4363 flux by fitting the blended lines with a single Gaussian profile and subtracting from its flux the \Hg\ flux obtained in a fit without \OIII$\lambda$4363 included. This method resulted in $F_{\left[OIII\right]\lambda4363} \geq 9.1\times10^{-20}$~erg~s$^{-1}$~cm$^{-2}$. Repeating the former analysis gives $T_e \geq 15000$~K and $12 + \log{\left(O/H\right)} \leq 7.72$. A lower limit on metallicity was derived by fitting the blended feature with \OIII$\lambda$4363 only and results in $T_e \leq 34000$~K and $12 + \log{\left(O/H\right)} \geq 7.08$. These limits are consistent with the best-fit value and place our source somewhat below the mass-metallicity relation at similar redshifts (Fig.\ref{fig:mass_metal}).

As a final check on these results we fit the blended \OIII$\lambda$4363 and \Hg\ feature by fixing the ratio of \Hbs and \Hg\ to the appropriate Balmer decrement and find $12 + \log{\left(O/H\right)} = 7.25_{-0.17}^{+0.20}$, which is completely consistent with the above estimates.

\subsection{Morphological/photometric fitting and stellar population properties}
 To constrain the stellar mass and star formation rate of the host galaxy we employ fractional spectral energy distribution (SED) fitting. To do this we employ the tool \texttt{ForcePho} (Johnson+, in prep), which enables us to forward model the light distribution using a combination of S\'ersic profiles. We perform spatially resolved photometry with \texttt{ForcePho} following the methodology detailed in \citesup{Baker2023,Tacchella2023}. In short we model the AGN and host galaxy as a central point source component and underlying host galaxy component respectively and fit the light distribution in the individual exposures of all 10 NIRCam bands simultaneously. This enables us to obtain accurate spatially resolved fluxes and morphological parameters for the galaxy. This approach has been used previously in \citem{Maiolino_AGN} \citesup{Baker2023, Tacchella2023}.

\begin{figure}
    \centering
    \includegraphics[width=1.0\textwidth]{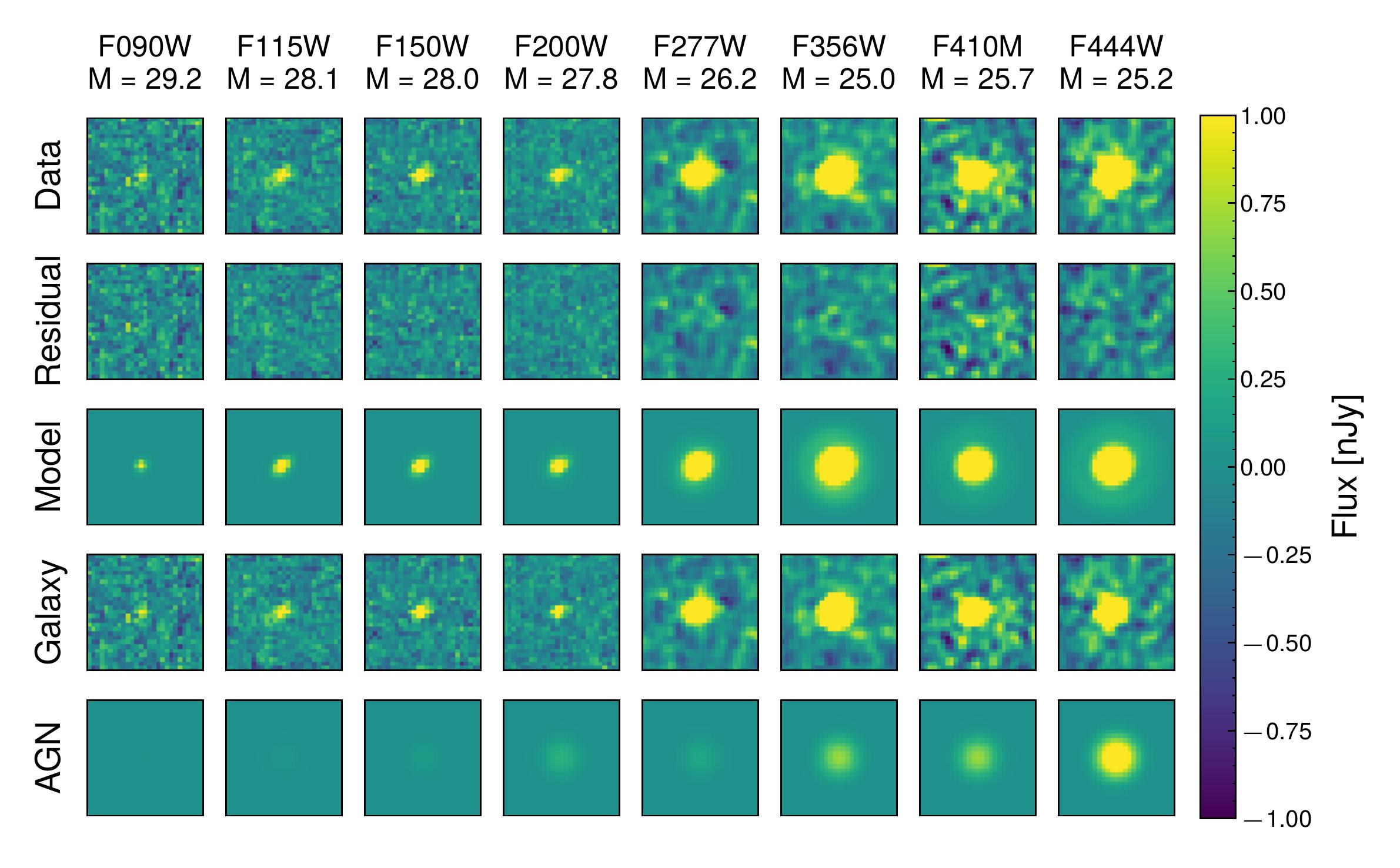}
    \caption{\textbf{AGN-host decompositions.} The data, residual, model, and fluxes with the recovered galaxy component, for the point source and galaxy decomposition in the 8 JADES NIRCam bands. The figure shows that the galaxy+point source model has fit the data well within all bands without leaving significant residuals. The bottom row shows the modeled point source component, stacked magnitudes in each band are shown above each column. Each panel is 0.8 by 0.8 arcsec in size.}
    \label{fig:forcepho_stack}
\end{figure}

 Fig.\ref{fig:forcepho_stack} shows the data, residual, model, and point-source subtracted host galaxy for the ForcePho fit. We can see that the galaxy + point-source model has fit the data well without leaving significant residuals and the galaxy component appears bright enough for reliable photometry, with S/N ranging from 6 - 40 across our filters. The resulting, best-fit morphological parameters are  reported in Table \ref{table:source_properties} and illustrate that the host galaxy is compact ($R_e = 137 \pm 8~{\rm pc}$) with a disk-like profile (Sersic index $n\sim 1$). The quoted statistical-only error on $R_e$ is rather small, considering the marginally resolved nature of the source.

\begin{figure}
    \centering
    \includegraphics[width=0.8\textwidth]{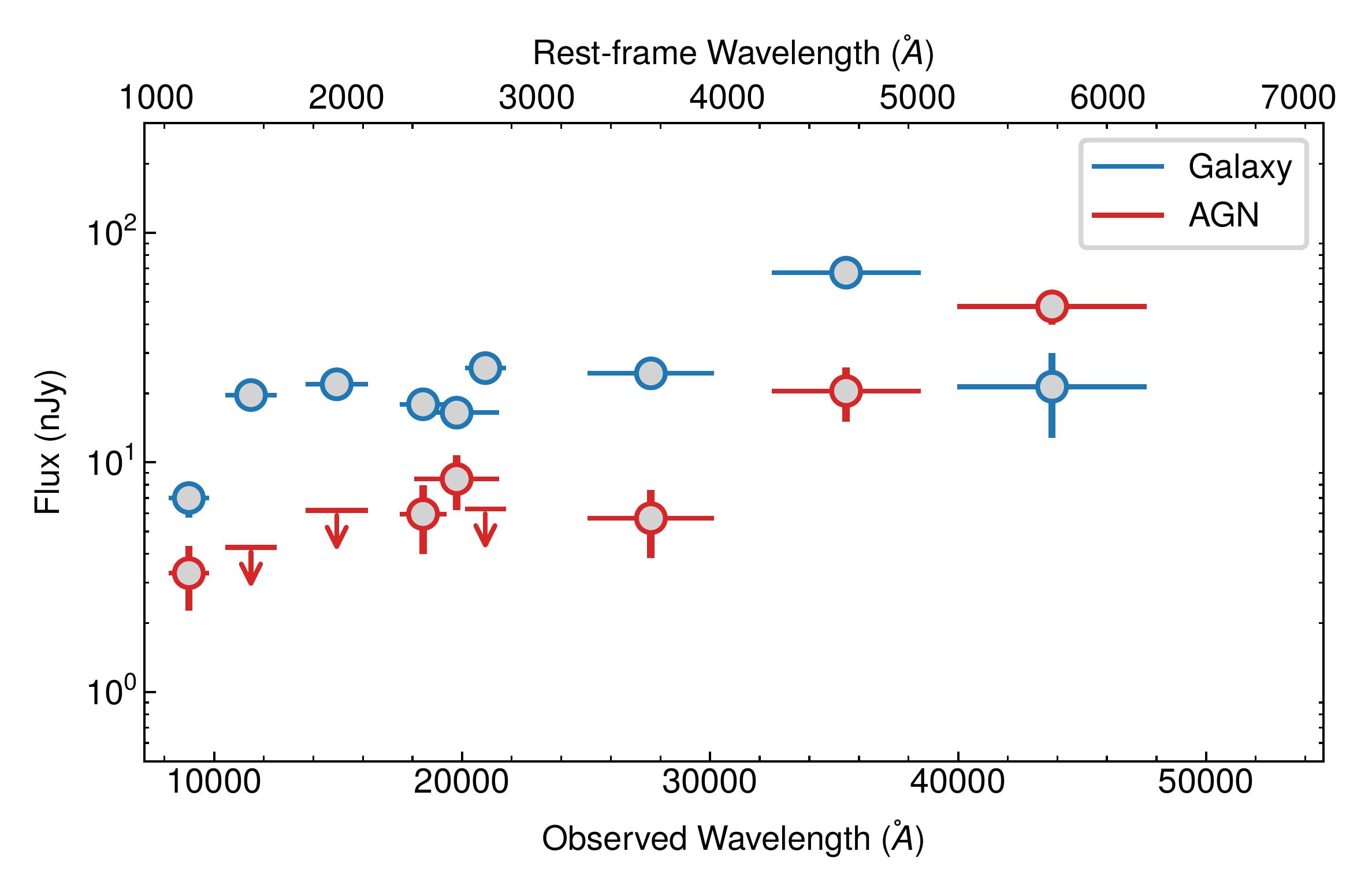}
    \caption{\textbf{AGN-host SEDs} The spectral energy distribution for the point source and galaxy decomposition in the 10 NIRCam bands. The figure shows that we have recovered a significant amount of the host galaxy's flux within all bands and that the AGN SED is reddened.}
    \label{fig:sed_comp}
\end{figure}

  The PSF model that is approximated by \texttt{ForcePho} is based off of WebbPSF - incorporating forward modelling of the telescope's optics, with additional calibration provided by field stars. To investigate the uncertainty coming from the PSF approximation used by \texttt{ForcePho}, we also re-fit the data with a different PSF approximation, that includes charge transfer effects, and obtain a 16\% smaller radius. We thus adopt a 16\% systematic error floor which results in a final estimate of $R_e = 137 \pm 23~{\rm pc}$. The fluxes for the point-source component and host galaxy can be seen in Fig.\ref{fig:sed_comp}. We find that the galaxy component dominated (at the 90\% level) in all fiters except for the F444W, where the contribution from the AGN broad H$\alpha$ dominates. 

 The next stage is to fit this spectral energy distribution in order to obtain the stellar population properties of the host galaxy. To do this we use the Bayesian SED fitting code \texttt{Prospector} \citem{Prospector}, which uses Flexible Stellar Population Synthesis \citesup[FSPS]{Conroy2009} with MIST isochrones \citesup{Choi2016}, nebular line and continuum emission \citesup{Byler2017} and a Chabrier \citesup{Chabrier2003} initial mass function (IMF). We also use the Bayesian SED fitting code \texttt{BAGPIPES} \citem{Carnall2018}, which uses stellar population models of Bruzual \& Charlot \citesup{BC2003}, alongside nebular line and continuum emission \citesup{Byler2017} with a Kroupa \citesup{Kroupa2001} IMF. 
 For both codes, we include a flexible two component dust model following \citesup{CharlotandFall2000} consisting of a birth cloud component (affecting only light from the the birth clouds themselves, e.g. stars younger than 10~Myr) and a separate `diffuse' component (affecting light from all sources). We assume a flexible star-formation history (SFH) with a continuity prior \citesup{Leja2019}, where we fit for the ratio between the six SFH bins. Finally, we exclude the F410M and F356W filters from the fit as those contain the \OIII$\lambda$$\lambda$5007,4959 doublet and thus may have been contaminated by AGN ionization. This can been seen in the SED in Fig.\ref{fig:sed_comp}, where we see the host galaxy has increased flux in the F410M and F356W bands compared to the point-source despite these bands containing \OIII$\lambda$5007$\lambda$4959. This suggests that emission line flux from the AGN may likely still be contributing the recovered host galaxy SED in these bands justifying our exclusion of these bands from the SED modelling.

\begin{figure}
    \centering
    \includegraphics[width=1\textwidth]{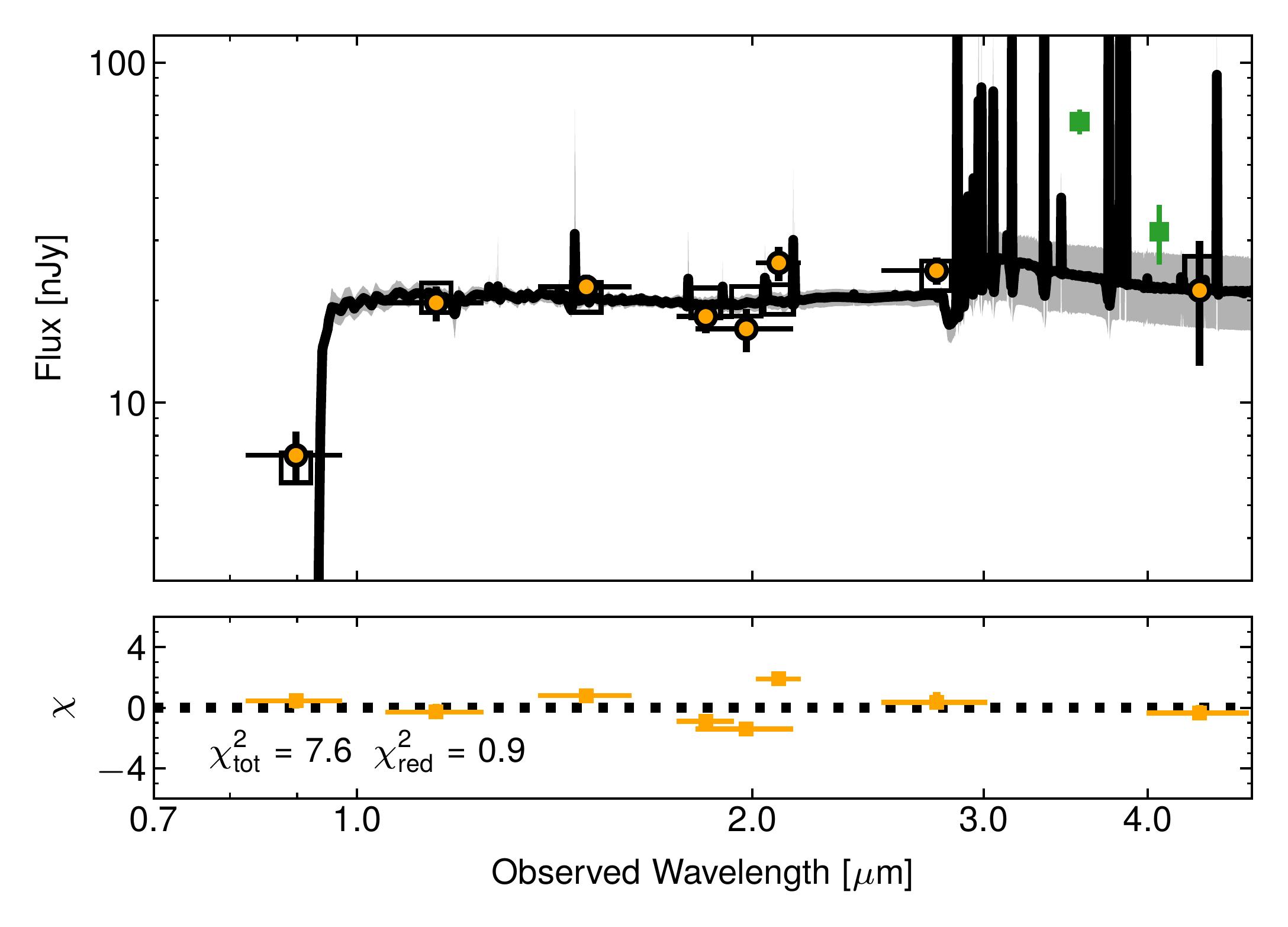}
    \caption{\textbf{Spectral Energy Distribution (SED) for the host galaxy fit by \texttt{Prospector}}. Yellow points correspond to the observed photometry. Black squares correspond to the model photometry and the model spectrum is overplotted in black. The chi distribution of the observed to model photometry is shown below. We note that we do not fit the F356W and F410M bands, shown in green, due to strong contamination from the AGN which is readily apparent in their excess flux relative to the other bands. The figure shows that \texttt{Prospector} has fit the observed photometry well.}
    \label{fig:sed_fit}
\end{figure}

 Fig.\ref{fig:sed_fit} shows the resulting \texttt{Prospector} fit (black) to the observed photometry (yellow) with the $\chi$ values below.
 These fits yield a stellar mass of $\log{(M_*/M_\odot)} = 9.00^{+0.27}_{-0.25}$ and $\log{(M_*/M_\odot)} = 8.71^{+0.24}_{-0.52}$ and instantaneous (within the last 10~Myr) star formation rates of $\rm SFR = 1.48^{+0.95}_{-0.42}\ M_\odot/yr$ and $\rm SFR = 1.25^{+1.13}_{-0.89}\ M_\odot/yr$ for \texttt{BAGPIPES} and \texttt{Prospector}, respectively. These results are consistent within 1$\sigma$ thus we combine the chains given by each code for the final estimate - $\log{(M_*/M_\odot)} = 8.92^{+0.30}_{-0.31}$ and $\rm SFR = 1.38^{+0.92}_{-0.45}\ M_\odot/yr$.

We also test the effect of fitting the SED of the combined photometry, i.e. host galaxy + point-source, to estimate how much we would overestimate the stellar mass of the galaxy by not decomposing the AGN and host galaxy components. We find that the results are altered by less than 1$\sigma$.

In order to further test the validity of our decompositions, we perform morphological analysis on the stacked NIRCam images of the source, comparing them to a model PSF. This comparison is carried out in two NIRCam bands - the F115W and F277W. The former was chosen due to its PSF being smallest among the bands not contaminated by the Lyman break, which prevents the use of F090W. The F277W band was chosen as it smears the source over more pixels, \textbf{better sampling the PSF}, while our \texttt{ForcePho} fits indicate that the PSF component is still  sub-dominant even in this filter.

\begin{figure}
    \centering
    \includegraphics[width=\linewidth]{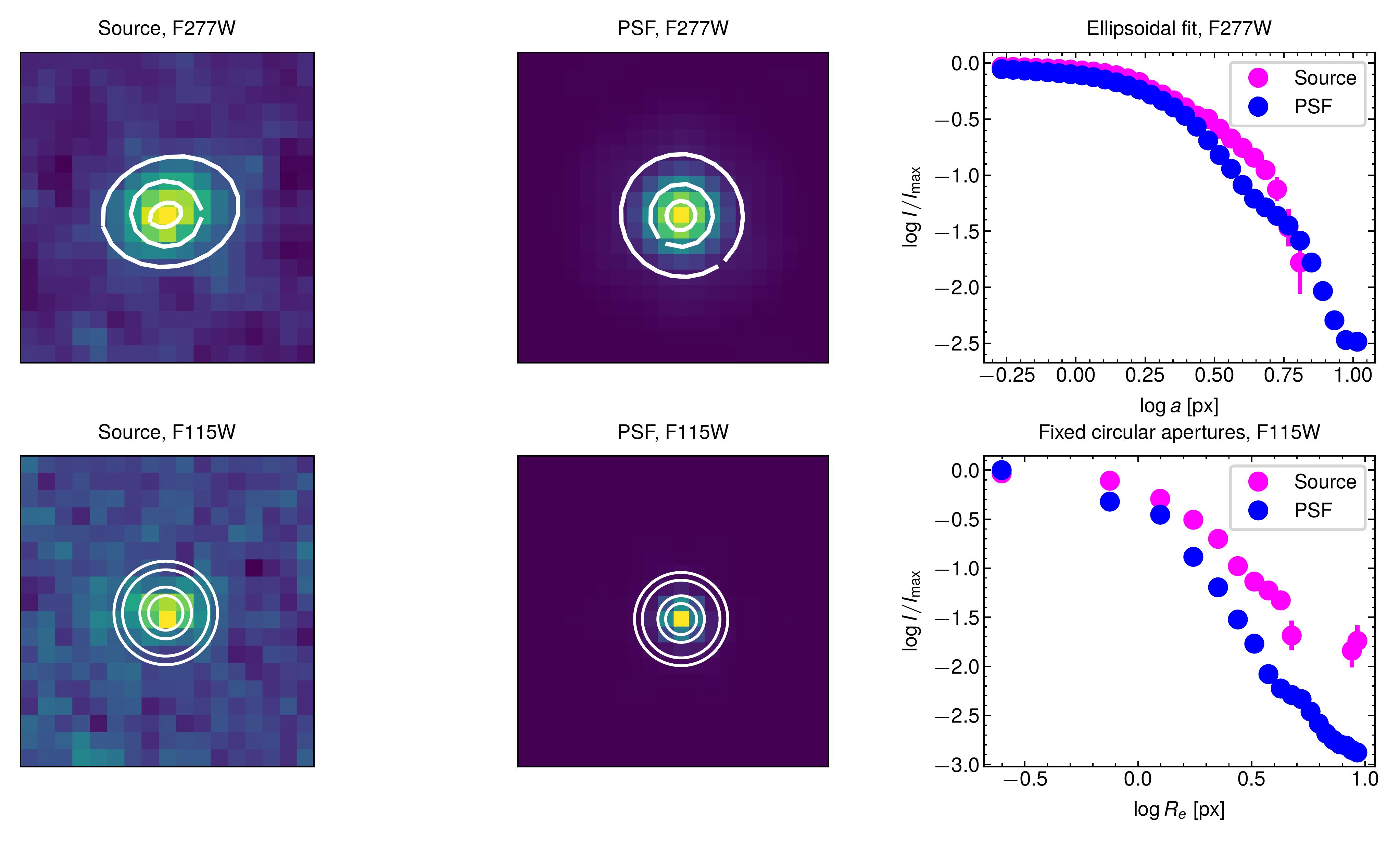}
    \caption{\textbf{Radial profile diagrams.}Comparisons of our object images with the model PSFs in the F277W (top row) and F277W (bottom row) bands. The top row shows the radial profiles for both source (magenta) and PSF (blue) obtained by fitting elliptical isophotes (shown in white in the cutouts). The bottom row shows the radial profiles in F115W band given by measuring fluxes in concentric circular apertures. All radial profiles measured are normalized to the maximum intensity and presented in log-scale. Both methods also reveal GN-1001830 to have a significant extended component.}
    \label{fig:radial}
\end{figure}

When analyzing the F277W band, we perform an isophotal fit for both the source and the model PSF using the \texttt{Photutils} package \citesup{photutils}. The results of this fit are shown in the top row of Fig.\ref{fig:radial}. As it can be seen there, the best-fit ellipsoidal isophotes for GN-1001830 are slightly elongated, suggesting some extended morphology, while the isophotes of the PSF are circular. The radial profiles, shown in the rightmost column of Fig.\ref{fig:radial}, clearly show that our object is more extended than the PSF.

Radial profiles in the F115W band were estimated by placing concentric circular apertures on both our source and the model PSF as the flux was dispersed over too few pixels, \textbf{undersampling the PSF and} making isophotal fits non-viable. Nevertheless, as shown in the bottom right of Fig.\ref{fig:radial}, our source, while compact,is still significantly more extended than the PSF.

All numeric properties of the host galaxy and its black hole obtained from the spectral and SED fitting are summarized in Extended Data Table \ref{table:source_properties}. 

\begin{table}
\centering
\label{table:source_properties}
\begin{tabular}{cc}
\toprule
Quantity  & Value \\[+8pt]

\midrule
$q$ & $0.51_{-0.04}^{+0.04}$ \\ [+5pt]
$n$ & $0.94_{-0.07}^{+0.07}$ \\ [+5pt]
$R_e$~[pc] & $137_{-23}^{+23}$ \\ [+5pt]
$\sigma_*$ [km~s$^{-1}$] & $121_{-6}^{+6}$ \\ [+5pt]
log($M_{\rm dyn}/M_\odot$) & $9.50^{+0.39}_{-0.39}$  \\ [+5pt]
SFR [$M_\odot/yr$] & $1.38^{+0.92}_{-0.45}$  \\ [+5pt]
log($M_*/M_\odot$) & $8.92^{+0.30}_{-0.31}$  \\ [+5pt]
$\log{M_{\rm BH}/M_{\odot}}$ & $8.61^{+0.38}_{-0.37}$  \\ [+5pt]
$\lambda_{\rm EDD}$ & $0.024^{+0.011}_{-0.008}$  \\ [+5pt]
$12 + \log{\left(O/H\right)}$ & $7.32_{-0.10}^{+0.16}$ \\ [+5pt]
$T_e$~[K]& $25400^{+3200}_{-4800}$ \\ [+5pt]
$A_V$ & $2.00_{-0.41}^{+0.44}$ \\

\bottomrule
\end{tabular}

\caption{\textbf{Properties of the host galaxy and its AGN obtained via imaging modelling, SED fitting and spectral analysis.}}

\end{table}

\begin{figure}
    \centering
    \includegraphics[width=\linewidth]{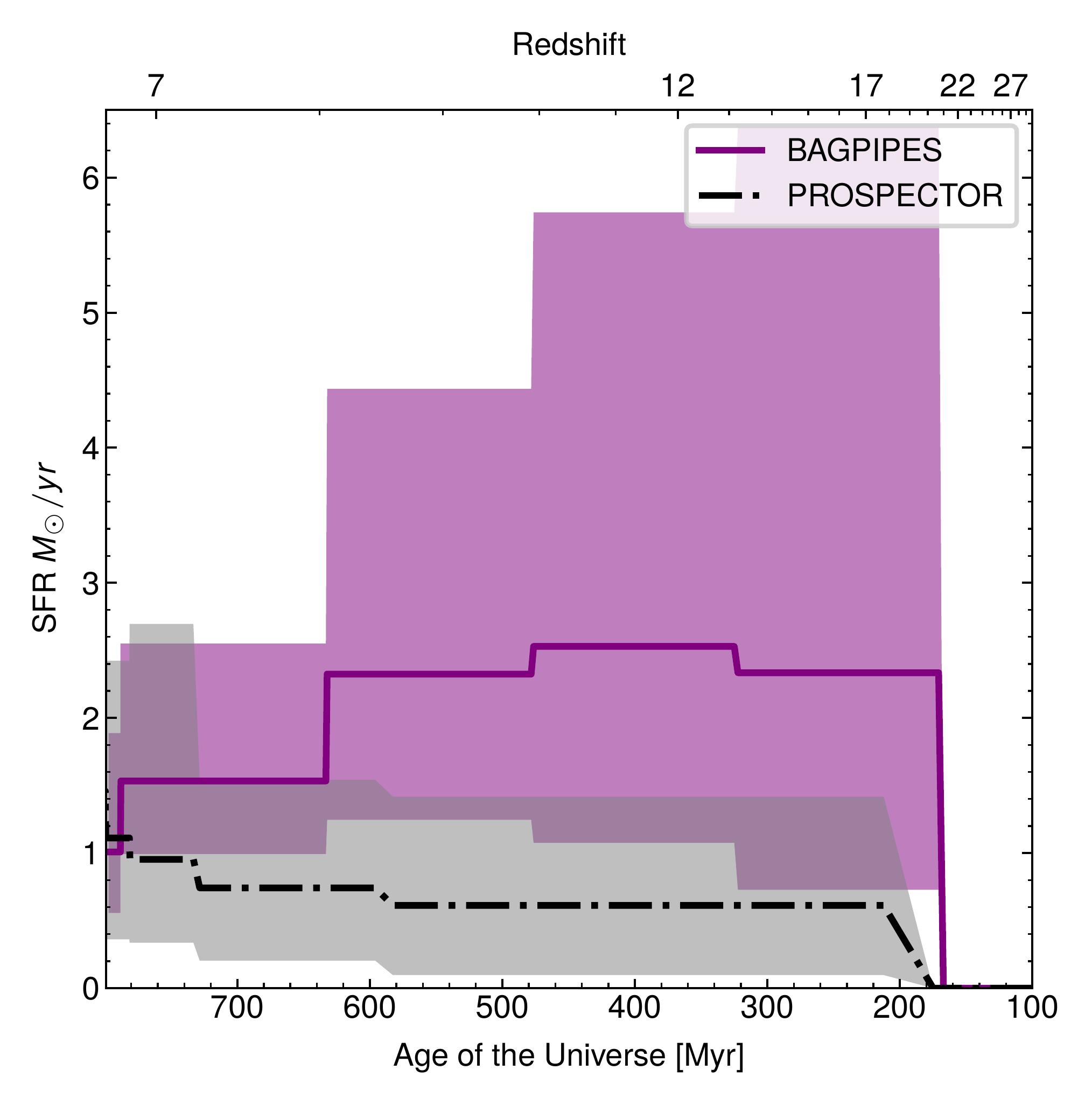}
    \caption{\textbf{SFH comparisons.} A comparison between the SFHs given by the two different codes used. While both SFHs are quite uncertain, they consistently show that the galaxy experienced a roughly constant SFR between 1 and 2 M$_{\odot}$~yr$^{-1}$ for most of its lifetime.}
    \label{fig:sfh}
\end{figure}

\begin{figure}
    \centering
    \includegraphics[width=1.0\textwidth]{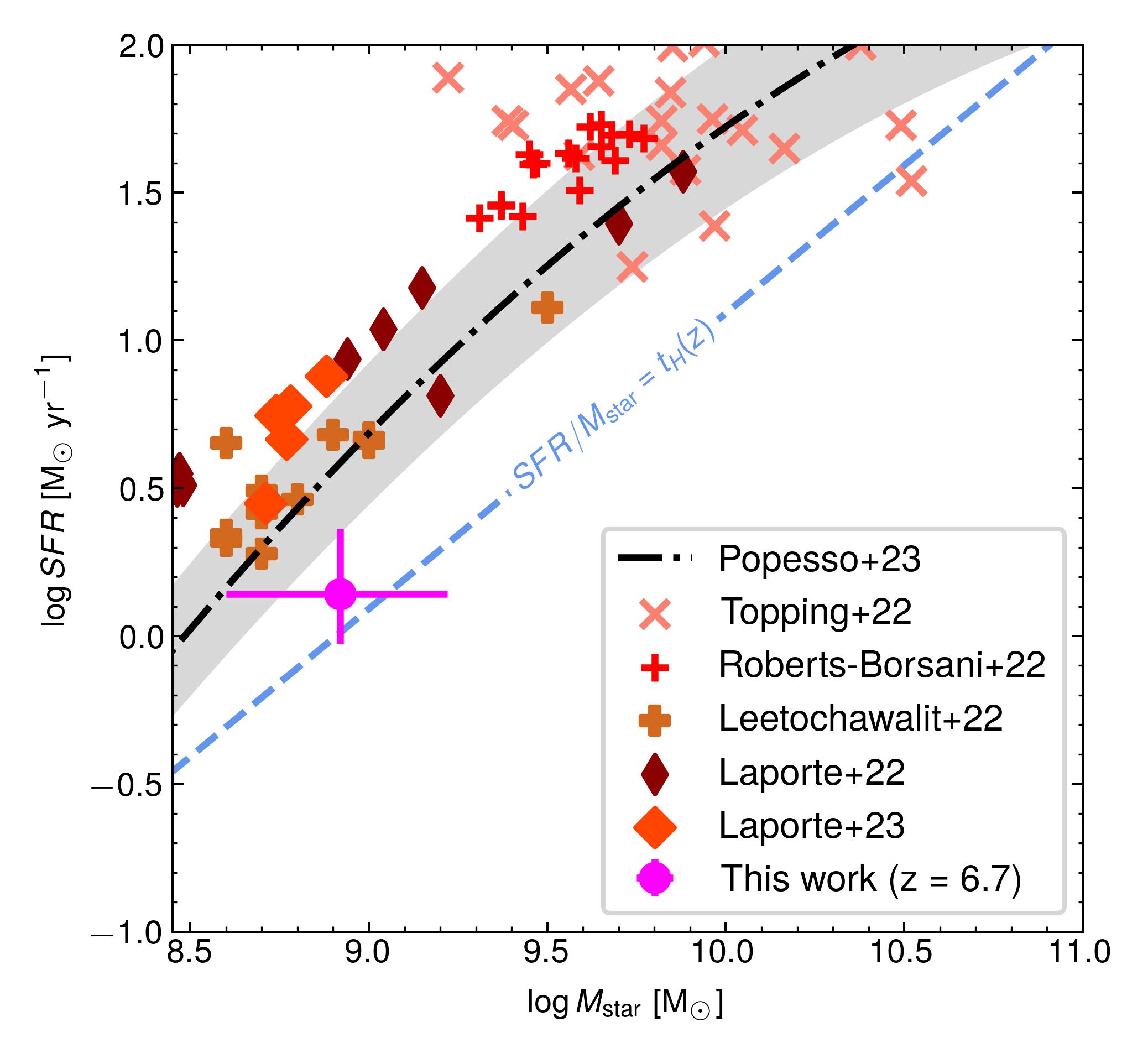}
    \caption{Star-formation rate versus stellar mass diagram showing the location of GN-1146115 with respect to the star forming main sequence. The black dash-dotted line shows the star forming main sequence fit from \protect\citesup{Popesso2023} extrapolated to z = 6.677, with the grey shaded area representing the uncertainties. Data at 7 $<$ z $<$ 9 from \protect\citesup{Laporte2023, Leethochawalit2023,  Laporte2022, Roberts-Borsani2022} and \protect\citesup{Topping2022} is shown with brown and red symbols. The dashed  blue line indicates the limit below which it takes a galaxy more than the Hubble time at z = 6.677 to double its mass. The magenta circle shows the location of GN-1146115, which is consistent with the dashed line within 1$\sigma$.}
    \label{fig:SFH_MS}
\end{figure}

Fig.\ref{fig:sfh} shows the SFR of the host galaxy against lookback-time (and redshift) presenting the star-formation history as derived by both \texttt{Prospector} and \texttt{Bagpipes}. Both SFHs are consistent with a flat SFH although the two codes show systematic offsets, likely resulting from different assumptions in the codes, in particular the different stellar populations used. This suggests that the galaxy experienced almost constant star formation over the last few 100 Myr. A comparison of GN-1001830 with the star forming main sequence at similar redshifts is shown in Fig.\ref{fig:SFH_MS}. Our source lies below the star-forming locus of similarly massive galaxies by a factor of $\sim$3, and would take $\sim$1~Gyr (i.e. about the age of the Universe at z$\sim$6) to double its mass with the current SFR. This indicates that the galaxy is currently fairly quiescent and may have been so for quite some time, although uncertainties on the SFH are large. The presence of AGN activity might offer a possible explanation for this state of the host, suggesting that AGN negative feedback might be responsible for suppressing star formation.

  We also fit the extracted point source (AGN) component with a reddened powerlaw with a fixed slope of $\beta = -1.55$ ($F_\lambda \propto \lambda ^\beta$), corresponding to average slope of type 1, unobscured high-z quasars \citesup{VandenBerk2001}. Assuming the SMC extinction curve, the resulting $A_V$ is $2.68 \pm 1.00$, which, while poorly constrained, is fully consistent with the value derived from spectroscopy. The bolometric luminosity inferred from the fitted $\lambda L_{\lambda}$ at 5,100\AA,  with the bolometric correction from \citesup{Saccheo2023}, and  corrected for absorption, is $4.6^{+7.5}_{-3.2}\times 10^{44} {\rm erg~s^{-1}}$, which is highly uncertain, but consistent with the value obtained from the broad component of H$\alpha$.

\subsection{Estimation of the Black hole mass and accretion rate}

As mentioned in the main text the broad component of H$\alpha$ can
be used to infer the black hole mass by assuming the local scaling relations being valid at high redshift and, specifically,
from the equation \citem{Reines2013,VolonteriBHmass}:

\begin{equation}
\label{eq:virial_mass}
    \log{\frac{M_{\rm BH}}{M_{\odot}}} = 6.60 + 0.47\log{\left(\frac{L_{H\alpha}}{10^{42}\ {\rm ergs^{-1}}}\right)+2.06\log\left(\frac{{\rm FWHM}_{H\alpha}}{1000 {\rm kms^{-1}}}\right)}
\end{equation}


The best fitting values for the broad component of H$\alpha$ (Full Width Half Maximum of $5700_{-1100}^{+1700}$~km~s$^{-1}$ and flux of $27.3^{+4.1}_{-4.0}\times10^{-19}$~erg~s$^{-1}$~cm$^{-2}$)
 give $\log{M_{\rm BH}/M_{\odot}} = 8.23^{+0.38}_{-0.36}$, with the scatter on \autoref{eq:virial_mass} included in the uncertainties.  Coupled with the bolometric luminosity of $\rm 2\times 10^{44}~erg~s^{-1}$, estimated from the broad component of H$\alpha$ (following the scaling relation given by \citesup{SternLbol}), gives an Eddington ratio $\rm \lambda_{Edd} = 0.009_{-0.003}^{+0.005}$, with a systematic scatter of 0.5~dex. 

We note that a recent measurement of the black hole mass in a super-Eddington accreting quasar at z$\sim$2 has cast doubts on the validity of the UV virial relations for single-epoch black hole measurements \citem{Gravity24}. The same work, however, points out that when using H$\alpha$ to measure the black hole mass, the discrepancy is only a factor of 2.5. Moreover, the discrepancy has been ascribed to deviations of the BLR size in the super-Eddington regime, which is certainly not the case for GN-1001830. In summary, the black hole mass measurement inferred from the broad H$\alpha$ in GN-1001830 is reasonably solid.

It is difficult to estimate dust attenuation in this object. The lack of broad H$\beta$ does not provide strong constraints: even assuming a standard case-B recombination ratio of 2.8, the broad H$\alpha$ line flux implies that the broad H$\beta$ is not detectable. We therefore assume, as found for other AGN at high-z, that the bulk of the obscuration towards the Broad Line Region also affects the narrow components \citesup{Gilli_dust}. From the observed ratio of the narrow components of H$\alpha$ and H$\beta$ ($5.51_{-0.69}^{+0.86}$), and assuming an SMC extinction law (appropriate for high-z AGN, \citesup{Richards2003, Reichard2003}), we infer $A_V = 2.00_{-0.41}^{+0.44}$~mag. We also repeat the estimate using ratios of \Has and \Hbs lines to \Hg. This yields $A_V = 2.31_{-0.61}^{+0.81}$~mag $ 2.6_{-2.1}^{+2.6}$~mag respectively. These values are quite uncertain due to lower brightness of \Hg\ and its proximity to the \Hbs line, but remain consistent with the previous estimate. The extinction corrected black hole mass, bolometric luminosity and Eddington ratio are $\log{M_{\rm BH}/M_{\odot}} = 8.61^{+0.38}_{-0.37}$ (with the uncertainty including intrinsic scatter on the virial relation), $L_{\rm bol}= 10^{45}~{\rm erg~s^{-1}}$, and $\lambda_{\rm Edd} = 0.024^{+0.011}_{-0.008}$, respectively, with the same intrinsic scatter of 0.5~dex. The extinction correction is uncertain due to the use of the narrow lines, however, the extinction-corrected values of BH mass and Eddington ratio are still consistent with the uncorrected ones within 2$\sigma$, hence our conclusions are not significantly altered by the presence of dust. Additionally, in the next section we show that fitting the nuclear source detected by NIRCam with a dust-reddened AGN slope results into an extinction consistent with that inferred from the narrow lines. In the previous section we also infer the bolometric luminosity from the fitting of the nuclear SED and, although with large uncertainties, independently obtain a value consistent with that obtained from the broad component of H$\alpha$. As an additional check, we infer the bolometric AGN luminosity from the luminosity of the narrow \Hbs and \OIII$\lambda$5007 lines using the scaling relations from \citesup{Netzer2009}. We obtain $L_{bol, H\beta(N)} = 4.7^{+0.7}_{-0.7}\times 10^{44}$~erg~s$^{-1}$ and $L_{bol, [OIII]} = 4.9^{+0.1}_{-0.1}\times 10^{44}$~erg~s$^{-1}$, which, while lower than the broad \Has estimate, are still consistent with it once the 0.3-0.4~dex scatter on the calibrations is taken into account. Additionally, we utilize the $5100 \AA$ luminosity from the previous section to independently infer the BH mass, which results in $\log{M_{\rm BH}/M_{\odot}} = 8.1 \pm 0.8$, with the large error coming from the uncertainties in source decomposition (being the AGN light subdominant), reddening and intrinsic scatter on the virial relations. This value is consistent with estimates using broad \Ha, but the uncertainty makes it rather unconstraining.

We note that even without correcting for extinction, the resulting {\it lower limit} on  the black hole mass would still imply a BH-to-stellar mass ratio several 100 times above the local relation.

\subsection{Black hole scaling relations with $\sigma$ and dynamical mass}

As discussed in  \citem{Maiolino_AGN}, while high-z AGN are offset on the BH-stellar mass plane relative to the local relation, they are much closer to the local relation between black hole mass and stellar velocity dispersion $\sigma_*$ relation and to the local relation between black hole mass and host galaxy dynamical mass. Here we explore the location of GN-1001830 on the latter two scaling relations.

\begin{figure}
    \centering
    \includegraphics[width=\textwidth]{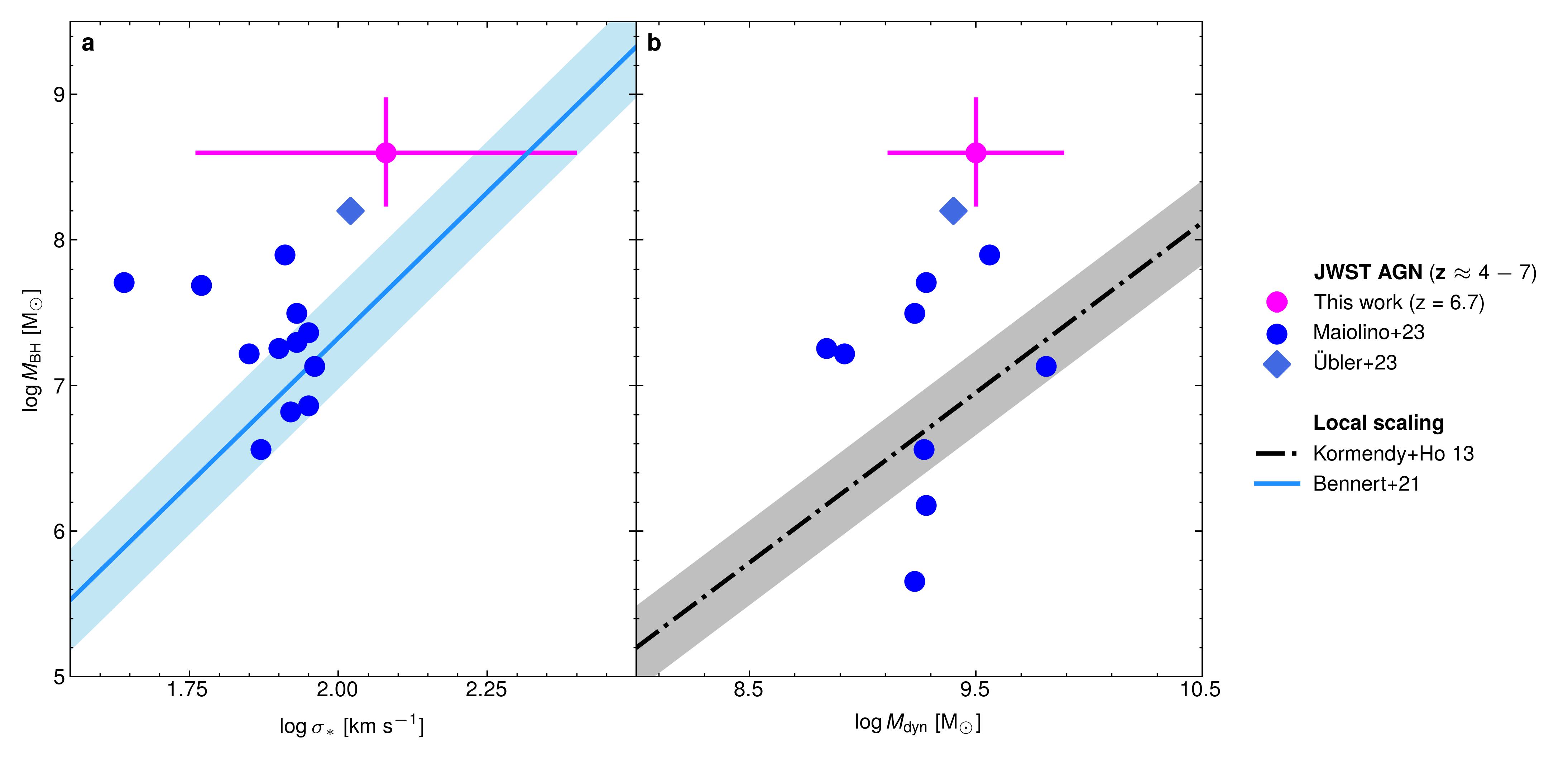}
    \caption{\textbf{Dynamical mass and velocity dispersion comparisons.} Location of GN-1001830 (magenta point) on the black hole mass versus stellar velocity dispersion (left) and versus dynamical mass of the host galaxy (right).  Other high-z AGN found by JWST are shown with blue symbols. The black dash dotted line shows the local $M_{\rm BH}$ - $M_{\rm bulge}$ relation from \protect\citesup{KormendyHo2013}. The solid blue line shows the $M_{\rm BH}$ - $\sigma_{*}$ relation from \protect\citesup{Bennert2021}. Shaded areas show the scatter around these relations. While not yet on the local relations, the offset of GN-1001830 is much less severe than in the BH-stellar mass diagram.}
    \label{fig:Mdyn_sigma}
\end{figure}

We use as a proxy of the velocity dispersion the width of the [OIII] line, $\rm FWHM=255\pm38~km~s^{-1}$, as measured from the medium resolution grating spectrum, deconvolved for the line spread function for compact sources. The uncertainty on the FWHM value for the narrow lines includes the 20\% systematic uncertainty of the LSF broadening, which is not significant for the broad component. We then derive the stellar velocity dispersion by correcting gaseous velocity dispersion by 0.1 dex to obtain the stellar velocity dispersion, following \citesup{Bezanson2018}, giving $\sigma_{*} = 121^{+16}_{-16}$~km~s$^{-1}$, with 0.3~dex intrinsic scatter for $\log{\sigma_*} = 2.08 \pm 0.32$. The resulting location of GN-1001830 in Fig.\ref{fig:Mdyn_sigma}, shown with a magenta circle, illustrates that our object is close to the local relation (solid line, with dispersion shown with a shaded region), as are other high-z AGN previously measured by JWST.

 We also utilize the host galaxy parameters to estimate its dynamical mass using the same approach as in \citem{Ubler2023, Maiolino_AGN}, which makes use of the equation:

\begin{equation}
\label{eq:m_dyn}
    M_{\rm dyn} = K(n)K(q)\frac{\sigma_{*}^{2}R_e}{G},
\end{equation}

where $K(n) = 8.87 - 0.831n + 0.0241n^2$, with S\'ersic index $n$, $K(q) = \left[0.87 + 0.38e^{-3.71(1-q)} \right]^{2}$, with $q$ being the axis ratio \citesup{Wel2022}, $R_e$ - the estimated effective radius and $\sigma_{*}$ - the stellar velocity dispersion. 
Utilizing \autoref{eq:m_dyn} along with values for $R_e$, $n$ and $q$ given by \texttt{ForcePho} (summarized in Extended Data Table \ref{table:source_properties}) gives $\log{M_{\rm dyn}/M_{\odot}} = 9.50^{+0.39}_{-0.39}$, with the uncertainty dominated by the intrinsic scatter on \autoref{eq:m_dyn}. We caution that the errors on this estimate are likely underestimated due to the absence of high resolution spectral observations for GN-1001830. In addition, the $M_{\rm dyn}$ value may be underestimated as our object is not fully centered in the slit (Fig. \ref{fig:rgb}), which may cut off part of the rotation curve in case of source rotation. The position of our object with respect to other JWST sources and the local scaling relation on the $M_{\rm BH}$ - $M_{\rm dyn}$ and $M_{\rm BH}$ - $\log{\sigma_*}$ star plane is shown in Fig.\ref{fig:Mdyn_sigma}. The source remains above the local relation, however, the difference is not as severe as in the BH - stellar mass relation.

\subsection{Gas fraction and star formation efficiency}

\begin{figure}
    \centering
    \includegraphics[width=\linewidth]{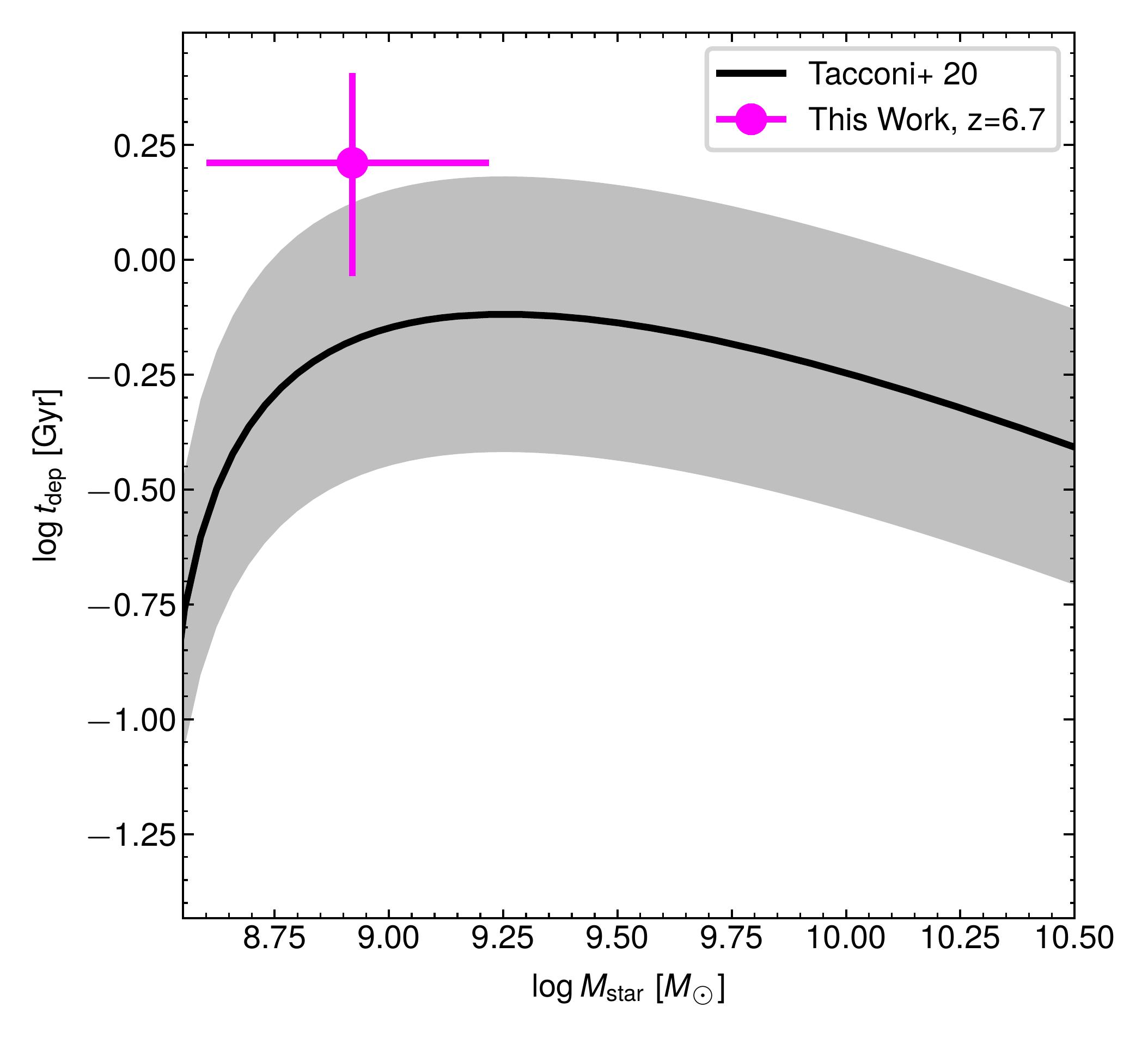}
    \caption{\textbf{Depletion time.} Location of GN-1001830 (magenta point) with respect to the depletion time versus stellar mass scaling relation for sources with the same SFR taken from \cite{Tacconi2020} (solid black line). The grey shading represents the $\sim$0.3~dex intrinsic scatter on the relation.}
    \label{fig:dep}
\end{figure}

Following our dynamical and stellar mass estimates we are able to obtain an estimate for the gas mass in the host galaxy, using $M_{\rm gas} = M_{\rm dyn} - M_{\rm star}$, assuming little contribution from dark matter, especially at such early epochs, within the central few 100 pc. This gives $\log{M_{\rm gas}/M_{\odot}} = 9.37^{+0.39}_{-0.39}$ and a gas fraction $f_{\rm gas}$ of $0.74^{+0.18}_{-0.30}$. The depletion time for our object can thus be estimated as $\frac{M_{\rm gas}}{\rm SFR} = 1.59_{-0.85}^{+0.91}$~Gyr. The depletion time given by the scaling relation from \citesup{Tacconi2020}, assuming a star forming main sequence of the form given in \citesup{Popesso2023}, evaluates to 0.66~Gyr. While these values are consistent within the intrinsic scatter on the relation, as shown in Fig.\ref{fig:dep}, this is still suggestive of star formation being inhibited relative to the population of normal star forming galaxies at this epoch.

\subsection{Completeness analysis}
\label{sec:comp}
In order to quantify the selection bias affecting the identification of AGN with broad emission lines, we run the fitting procedure described in \ref{sub:spec_fit}  on simulated broad \Has profiles. These profiles were simulated by using the error extension of the prism data for GN-1001830 to simulate Gaussian noise and adding Gaussian line profiles along with a power-law continuum on top of it.  The FWHM and luminosity for the narrow \Has component were uniformly sampled from $\rm FWHM \in [200, 600]$~km~s$^{-1}$ and $\log{\left(L/{\rm erg\ s^{-1}}\right)} \in [42, 43]$ ranges respectively. Redshifts of the simulated sources were uniformly drawn from between 6 and 7. Continuum normalization was set to be 100 times smaller than the \Has narrow peak height and slopes were sampled from a uniform distribution  ($-1 < \alpha < 1$). The model grid for the broad component of \Has was computed by varying $\log{\lambda_{\rm Edd}}$ between -3.0 and 0.5 in steps of 0.25 and the $\log{M_{\rm BH}/M_{\odot}}$ between 5.5 and 9 in steps of 0.25. For bins with $\log{M_{\rm BH}/M_{\odot}} < 7$ we performed the simulation assuming the grating spectrum R1000 error extension to assuage the effects of prism's lower resolution. \autoref{eq:virial_mass} and scaling relations from \citesup{SternLbol} were used to convert Eddington ratio and BH mass values to the luminosity and width of the simulated broad component. This yielded a 15x15 grid, each point of which contained 100 spectra simulated according to the above recipe.

The fitting of the simulated data was carried out with the same parameters as those described in \ref{sub:spec_fit}, with a bounded least-squares procedure being used to make the fitting computationally tractable. The completeness of each grid point was calculated as the ratio of the sources recovered according to our criteria (in terms of both broad line significance and $\Delta BIC$) to the sources inserted. The final completeness function is presented in Fig.\ref{fig:completeness} and shows that we are inherently biased against highly sub-Eddington black holes predicted by simulations, as most of those lie in regions of low completeness. We caution that the procedure described here only provides completeness with respect to the final step of AGN selection - the fitting of the BLR - and only covers the medium-depth tier of the JADES survey. The full selection function for the JADES survey is more complex as it spans multiple survey tiers, telescope instruments and source selection methods. Its full treatment is thus beyond the scope of the paper and our results here should be regarded as closer to an upper limit.

\subsection{The role of selection effects on the $M_{BH}-M_{star}$ relation}

As discussed in the main text, the finding that most of the black holes at high-z newly discovered by JWST are overmassive on the $M_{BH}-M_{star}$ could be due to a selection effect, i.e. the scatter of the relation is much larger at high redshift and more massive black holes tend to be preferentially selected at high redshift because they are, on average, more luminous.
We have shown and confirmed that selection effects do play an important role; however, our findings indicate that they cannot completely explain the offset relative to the local $M_{BH}-M_{star}$ relation.

Specifically, the scenario proposed by \citem{Li2024} (which envisages a BH-stellar relation similar to the local one, but with an order of magnitude scatter) hardly reaches the BH-to-stellar mass ratio observed in our object, and their observational bias scenario would require an observed luminosity of $\sim 10^{45-46}~{\rm erg/s}$, i.e. 1-2 dex above the luminosity observed in GN-1001830 prior to dust obscuration correction (we note that they do not take into account dust exinction, while this and most JWST-discovered AGN are affected by extinction).

The Trinity simulation \citem{TrinityIV} can produce overmassive black holes as observed in GN-1001830, but they require even higher luminosities, in excess of $\rm 10^{48}~erg/s$, hence totally incompatible with the luminosity of our object.
Therefore, while selection effects are important (as we illustrate more thoroughly in \ref{sec:comp}), our finding suggests that the overmassive nature of high-z BHs is also associated with an intrinsic offset of the BH-stellar mass relation, as also suggested by other studies \citem{Pacucci2023}.

Finally, our finding that GN-1001830, as well as many other JWST-discovered AGN \citem{Maiolino_AGN}, is closer (or even consistent for many AGN) with the $M_{BH}-\sigma$ and $M_{BH}-M_{star}$ relations, indicate that the selection effects on BH mass cannot play a major role, or else the same strong, orders of magnitude, offset should also be present on these relations.

These aspects are however outside the scope of this paper and will be discussed more extensively in a dedicated paper.

\subsection{Comparison with the FABLE simulations}
In order to put our work in greater context we also compare the properties of our source to predictions from the FABLE (\emph{Feedback Acting on Baryons in Large-scale Environments}) \citesup{Henden2018} simulations. These are carried out with the massively parallel AREPO code \citesup{AREPO} with new comoving 100 $h^{-1}$ Mpc boxes.  We consider both Eddington-limited and super-Eddington accretion, bounding our black hole Bondi-Hoyle-Lyttleton accretion rate in two different simulations by 1 and 10$\times$ Eddington. Details regarding other subgrid models, which are largely based on the Illustris galaxy formation models \citesup{Vogelsberger2014}, can be found in \citesup{Henden2018}.

\begin{figure}
    \centering
    \includegraphics[width=\textwidth]{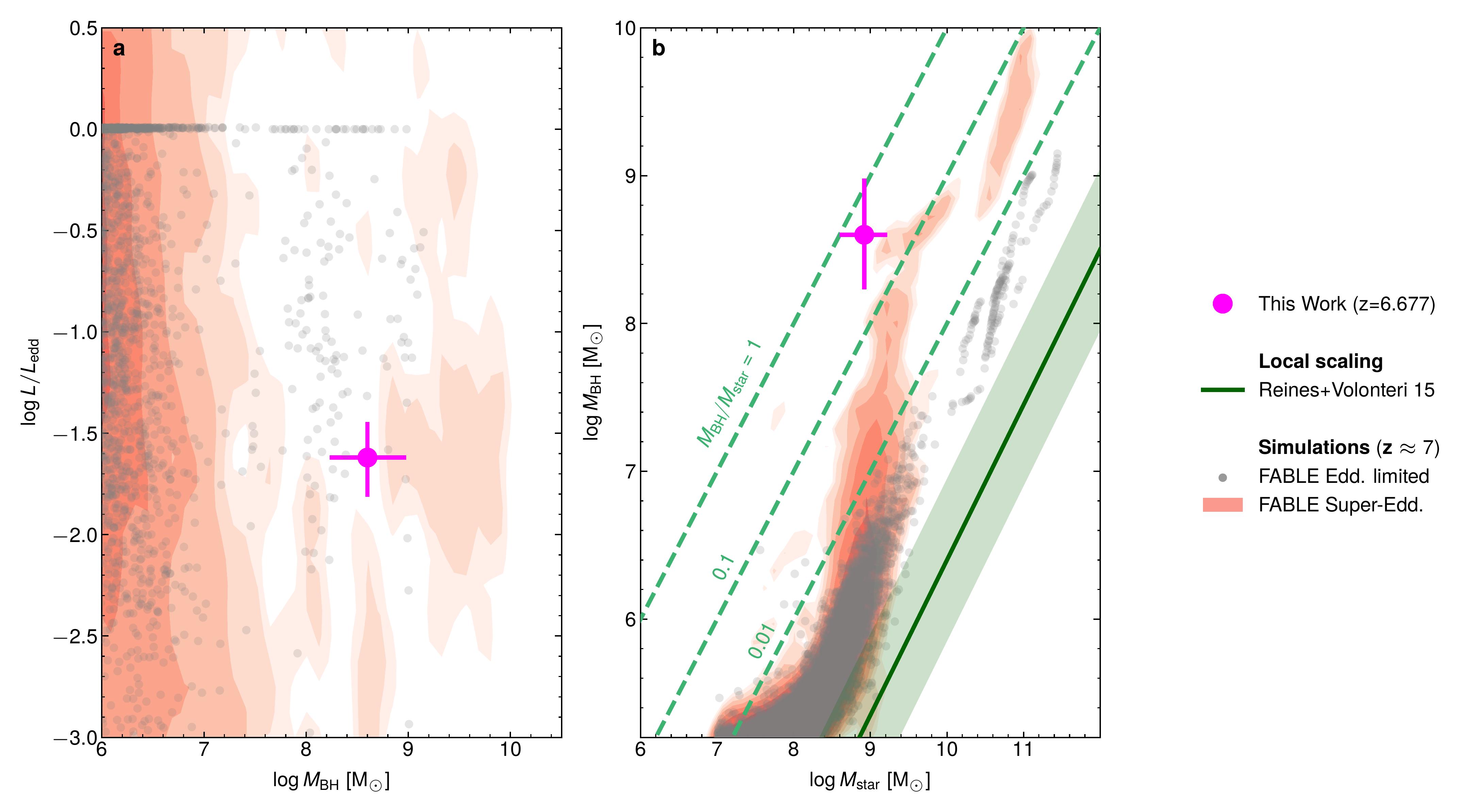}
    \caption{\textbf{Comparison between GN-1001830 and  sources in the FABLE simulation.} The layout is the same as the bottom row of Fig.\ref{fig:comparisons}, with sources from super-Eddington simulations shown as red contours, while those from sub-Eddington ones are indicated by grey points. As in the case of the CAT models, the Eddington-limited heavy seed scenario fails to simultaneously explain the high BH-to-stellar mass ratio of GN-1001830 and the very low accretion rate. Instead,  the scenario in which BHs experience super-Eddington accretion phases can match the properties of GN-1001830, although additional simulations would be need to bridge the gap between the 100 $h^{-1}$ Mpc box and proto-cluster zoom-in simulations.}
    \label{fig:fable_comp}
\end{figure}

A comparison between the simulated sources from FABLE and our object is provided in Fig.\ref{fig:fable_comp}, with the same coding as in Fig.\ref{fig:comparisons}, i.e. gray symbols show the Eddington-limited scenarios and the red contours the distribution of simulations in which super-Eddington accretion is allowed. As can be seen there, the super-Eddington simulation is able to produce more massive black holes with respect to both luminosity and stellar mass in comparison to the Eddington-limited simulation, resulting in a better match with our observations. However, GN-1001830 still lies above our super-Eddington simulations in the right-hand $M_{\rm BH}-M_{\rm star}$ plot, likely due to the FABLE simulations lacking in volume. To better sample a larger volume we also include simulation data from zoom-in simulations of a massive protocluster ($M_\mathrm{h} > 10^{12}$ M$_{\odot}$ at z=6.66), previously used in \citem{Bennett2024}, that is taken from the larger Millennium simulation volume of 500 $h^{-1}$ Mpc \citesup{Springel2005}. These points occupy the high-mass region in both panels of Fig.\ref{fig:fable_comp}. The objects obtained in these simulations have stellar masses slightly larger than inferred from GN-1001830 and, in runs including super-Eddington accretion up to 10$\times$ Eddington (and also earlier BH seeding, unlike the other FABLE results shown here, see \citem{Bennett2024}), have more massive black holes. Although a more quantitative match will be explored in future work, these results qualitatively show that super-Eddington bursts more readily explain the properties and presence of objects like GN-1001830, particularly in increasing the black hole to stellar mass ratio, relative to Eddington-limited scenarios.

\begin{figure}
    \centering
    \includegraphics[width=\textwidth]{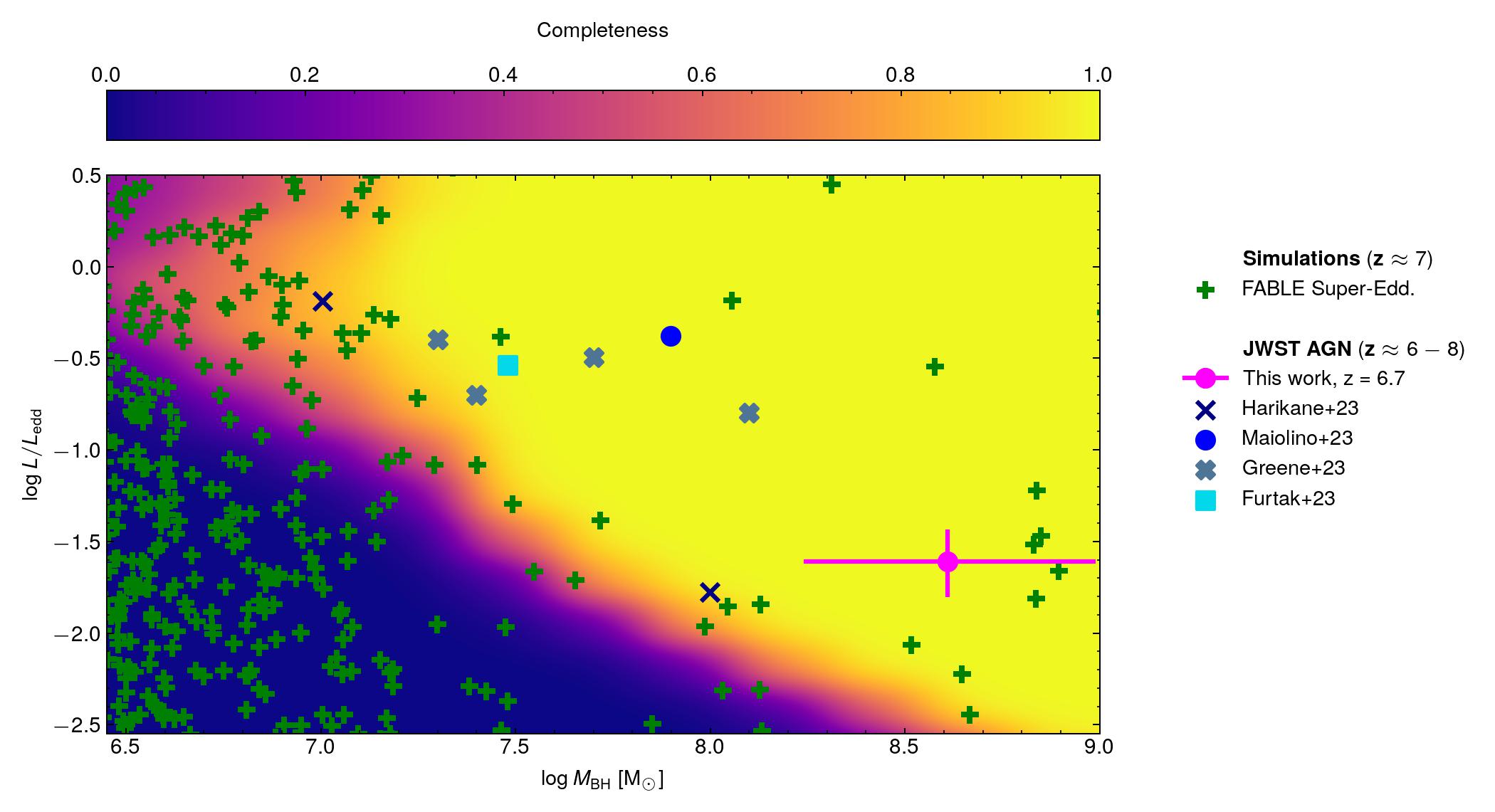}
    \caption{\textbf{Completeness simulations.} Same as Fig.\ref{fig:completeness} in the main text, except with super-Eddington models of the FABLE simulation, showing that dormant, post super-Eddington burst AGN are strongly biased against in the current surveys.}
    \label{fig:completeness_brian}
\end{figure}

We also note that, as with CAT sources discussed in the main text, most of the simulated highly sub-Eddington sources reside in the low completeness region of the JADES survey (Fig.\ref{fig:completeness_brian}) and are thus hard to detect at these redshifts even with current instruments.

\bibliographysup{sn-bibliography}


\end{document}